\documentclass[%
 reprint,
usenatbib,
 amsmath,amssymb,
 aps,
 nofootinbib, 
twocolumn]{aastex631}

\usepackage{graphicx}
\usepackage{dcolumn}
\usepackage{bm}
\usepackage{xcolor}
\usepackage{makecell} 
\usepackage{tabularx}
\usepackage{physics}
\usepackage{comment}
\usepackage{multirow}

\newcommand{\R}[0]{\mathbb{R}}
\newcommand{\C}[0]{\mathbb{C}}

\newcommand{\msun}{{\,\rm M_\odot}}

\newcommand{\kms}{\,{\rm km}\,{\rm s}^{-1}}
\newcommand{\cm}{\,{\rm cm}}

\newcommand{\K}{\,{\rm K}}

\newcommand{\pc}{\,{\rm pc}}
\newcommand{\kpc}{\,{\rm kpc}}

\newcommand{\me}{m_{\rm e}}
\newcommand{\mproton}{m_{\rm p}}
\newcommand{\rsprime}{r_s'}
\renewcommand{\arraystretch}{1.8}
\newcolumntype{C}[1]{>{\centering\let\newline\\\arraybackslash\hspace{0pt}}m{#1}}

\def\aap{A\&A}

\def\apjl{ApJ}

\def\apjs{ApJS}

\newcommand{\pprime}{p'}
\newcommand{\eprime}{e'}
\newcommand{\alphaprime}{\alpha'}
\newcommand{\meprime}{m_{e'}}
\newcommand{\mpprime}{m_{p'}}
\newcommand{\fprime}{{f'}}

\newcommand{\xiprime}{{\xi'}}

\newcommand{\Tcutoff}{T_{\rm cut-off}'}

\newcommand{\tcool}{t_{\rm{cool}}'}

\newcommand{\mvir}{M_{\rm{vir}}}
\newcommand{\rvir}{R_{\rm{vir}}}

\newcommand{\Tvir}{{T}_{\rm{vir}}'}

\newcommand{\lagrangian}{\mathcal{L}}

\newcommand{\Tcmb}{T_{\rm{cmb}}}

\newcommand{\Ebind}{E_{\rm{b}}'}
\newcommand{\Ebindeff}{E_{\rm{b,eff}}'}
\newcommand{\Ebindstd}{E_{\rm{b}}}

\newcommand{\thubble}{t_{\rm H}}
\newcommand{\betacool}{\beta_{\rm cool}'}

\newcommand{\alphae}{{\alpha_{\rm E}}}
\newcommand{\rhoadm}{\rho_{\rm adm}'}
\newcommand{\rhocdm}{\rho_{\rm cdm}}
\newcommand{\Mdm}{M_{\rm dm}}
\newcommand{\rhoadmtwiddle}{\Tilde{\rho}_{\rm adm}}
\newcommand{\rhocdmtwiddle}{\Tilde{\rho}_{\rm cdm}}
\newcommand{\deltaPoissonJeans}{\delta_{\rm PJ}^2}
\newcommand{\deltamass}{\delta_{\rm mass}^2}
\newcommand{\deltatotal}{\delta_{\rm total}^2}
\newcommand{\vcirc}{V_{\rm circ}}

\newcommand{\rhalf}{R_{1/2}}
\newcommand{\zhalf}{Z_{1/2}}

\newcommand{\rninety}{R_{9/10}}
\newcommand{\zninety}{Z_{9/10}}
\newcommand{\flatness}{\Tilde{\epsilon}}

\newcommand{\Bhightlow}{\texttt{superfast}}
\newcommand{\Bhighthigh}{\texttt{fast}}
\newcommand{\Bhighthighfhigh}{\texttt{fast-f12\%}}
\newcommand{\Bhightextreme}{\texttt{slow}}
\newcommand{\Blowtlow}{\texttt{superfast-Ebindlow}}
\newcommand{\Blowthigh}{\texttt{fast-Ebindlow}}
\newcommand{\Bhightlowmv}{\texttt{superfast-m10v}}
\newcommand{\mq}{\texttt{m10q}}
\newcommand{\mv}{\texttt{m10v}}

\newcommand{\rhoclump}{\rho_{\rm clump}'}

\newcommand{\Htwoprime}{\rm{H}'_2}

\newcommand{\Hneutralprime}{\rm{HI}'}

\newcommand{\SR}[1]{\textcolor{orange}{[SR: #1]}} 
\newcommand{\ML}[1]{\textcolor{green}{[ML: #1]}} 

\newcommand{\JB}[1]{\textcolor{red}{[JB: #1]}}
\newcommand{\phil}[1]{\textcolor{blue}{[PH: #1]}}
\newcommand{\DC}[1]{\textcolor{magenta}{[DC:#1]}}

\begin{document}

\title{Aggressively-Dissipative Dark Dwarfs: \\
The Effects of Atomic Dark Matter on the Inner Densities of Isolated Dwarf Galaxies}

\author[0000-0002-7638-7454]{Sandip Roy}
\affiliation{Department of Physics, Princeton University, Princeton, NJ 08544, USA}

\author[0000-0002-6196-823X]{Xuejian Shen}
\affiliation{Department of Physics,
Massachusetts Institute of Technology, Cambridge, MA 02139, USA}
\affiliation{Kavli Institute for Astrophysics and Space Research, Massachusetts Institute of Technology, Cambridge, MA 02139, USA}

\author[0000-0001-9043-6577]{Jared Barron}
\affiliation{C.N. Yang Institute for Theoretical Physics, Stony Brook University, Stony Brook, NY 11794-3800}

\author[0000-0002-8495-8659]{Mariangela Lisanti}
\affiliation{Department of Physics, Princeton University, Princeton, NJ 08544, USA}
\affiliation{Center for Computational Astrophysics, Flatiron Institute, New York, NY 10010, USA}

\author[0000-0003-0263-6195]{David Curtin}
\affiliation{Department of Physics, University of Toronto, Toronto, Ontario M5S 1A7, Canada}

\author[0000-0002-8659-3729]{Norman Murray}
\affiliation{Canadian Institute for Theoretical Astrophysics, University of Toronto, Toronto, ON M5S 3H8, Canada}

\author[0000-0003-3729-1684]{Philip F. Hopkins}
\affiliation{TAPIR, California Institute of Technology, Pasadena, CA, 91125, USA}

\begin{abstract}
We present the first suite of cosmological hydrodynamical zoom-in simulations of isolated dwarf galaxies for a dark sector that consists of Cold Dark Matter and a strongly-dissipative sub-component.  The simulations are implemented in GIZMO and include standard baryons following the FIRE-2 galaxy formation physics model. The dissipative dark matter is modeled as Atomic Dark Matter~(aDM), which forms a dark hydrogen gas that cools in direct analogy to the Standard Model. Our suite includes seven different simulations of $\sim 10^{10}\msun$ systems that vary over the aDM microphysics and the dwarf's evolutionary history.
We identify a region of aDM parameter space where the cooling rate is aggressive and the resulting halo density profile is  universal.  
In this regime, the aDM gas cools rapidly at high redshifts and only a small fraction survives in the form of a central dark gas disk; the majority collapses centrally into collisionless dark ``clumps'', which are clusters of sub-resolution dark compact objects.  These dark clumps rapidly equilibrate in the inner galaxy, resulting in an approximately isothermal distribution that can be modeled with a simple fitting function.  
Even when only a small fraction~($\sim 5\%$)  of the total dark matter is strongly dissipative, the central densities of classical dwarf galaxies can be enhanced by over an order of magnitude, providing a sharp prediction for observations. 
\end{abstract}

\section{Introduction}
\label{sec:introduction}

In the Cold Dark Matter~(CDM) paradigm, the dark matter interacts only through gravity.  However, many well-motivated theories suggest that the dark matter sector can undergo more complex interactions, just like particles in the Standard Model.
One example is the class of ``dark sector'' models that contain  cosmologically stable dark particles charged under new forces---see e.g.~\cite{Bertone:2018krk} for a review. These models include Hidden Valleys~\citep{Strassler:2006im}, such as the Twin Higgs model~\citep{Chacko:2005pe, Chacko:2016hvu, Chacko:2018vss} that solves the little hierarchy problem, 
and Mirror Dark Matter~\citep{Berezhiani:2005vv,
Foot:2004pa,
Mohapatra:2000qx}.  

A dark sector may contain ``dissipative'' dark matter particles that can lose energy through their self interactions.  This yields fundamentally different behavior to CDM on galactic scales, because the dark matter can cool and collapse, altering its overall distribution in a halo. If the degree of dissipation is strong enough, such models may  modify the mass function of galactic subhalos~\citep{Gurian:2021qhk} and/or lead to the formation of new substructures such as dark disks~\citep{Fan2013} and dark compact objects~\citep{Ghalsasi2017,Curtin:2019ngc,Gurian:2022nbx,Bramante2023,Fernandez2024}.  A detailed understanding of whether and how such substructures form  necessitates the use of cosmological hydrodynamical simulations, given the complexity of gravitational collapse processes, as well as the potential backreaction of the changing dark matter distribution on the galaxy's baryons.

To date, inelastic dark matter has been simulated for only a limited range of models. Effective multi-state dark matter models with endothermic and exothermic self interactions were studied in \citet{Medvedev2010a,Medvedev2010b,Medvedev2014a,Medvedev2014b,Schutz2015,Blennow2017,Zhang2017,Todoroki2019,Vogelsberger2019,oneil2023,Leonard2024}. Inelastic dark matter scattering models with fixed, fractional energy loss were also simulated in \citet{Shen2021,Xiao2021,Shen2024}. These studies assume that \emph{all} of the dark matter is \emph{weakly} self interacting and inelastic, which may lead to the  evaporation of dark matter from halos via exothermic collisions and the enhancement of inner halo densities via endothermic and dissipative collisions.

In this paper, we explore an alternative scenario where a {\em small fraction} of the dark matter is {\em strongly} self interacting and dissipative~\citep{McCullough2013,Fan2013,Fan2013c,Agrawal2016,Agrawal2017,Beauchesne2021,Yuan2024}. As a case study, we focus on minimal Atomic Dark Matter~(aDM), a useful benchmark given its analogous structure to ordinary baryons.  Minimal aDM consists of a fermionic dark proton, $p'$, and a fermionic dark electron, $e'$, that interact through a massless dark photon with coupling, $\alphaprime$~\citep{Kaplan2009}. The aDM can form a dark hydrogen bound state, $\rm HI'$, which radiatively cools, as well as dark molecular Hydrogen~($\rm H_2'$).
This minimal aDM model assumes no dark nuclear physics and that the dark sector only couples to the Standard Model through gravity. Moreover, the aDM abundance is set asymmetrically, so that the abundance of dark anti-particles is negligible~\citep{Zurek2013,Kaplan2011}.
%

aDM can have important effects on cosmological scales~\citep{cyrracine2012,cyr-racine2013}. In particular, its effects on large-scale structure and the Cosmic Microwave Background~(CMB) exclude significant portions of the parameter space, particularly large aDM mass fractions with late dark recombination, large dark sound horizons, and high dark CMB temperatures. All these effects impact well-quantified observables, such as the number of free-streaming species in the early Universe, the spectrum of temperature anisotropies in the CMB, and the clustering of galaxies on large scales~(see \citet{cyr-racine2013,Cyrracine2021,Bansal:2022qbi,Greene2024} for more details).

The remaining allowed parameter space can still have significant effects on sub-galactic scales, even when the aDM fraction is sub-dominant to CDM.  This was shown explicitly by  \citet{Roy:2023zar}, which presented the first simulations of Milky Way-mass galaxies in a CDM plus aDM cosmology.  These simulations illustrated for the first time how the aDM would cool in a galaxy like our own, forming a dark gas disk at its center. 
 The dark disk is unstable and fragments into dark ``clumps'' that are concentrated in the inner-most regions of the halo.  Physically, these clumps correspond to clusters of sub-resolution dark compact objects.  
 Additionally, \citet{Gemmell2024} demonstrated that the properties of the satellites in these Milky Way-mass systems may be altered from the CDM expectation, with differences in the distribution of their masses, concentrations, and locations.  

The results of both \citet{Roy:2023zar,Gemmell2024} demonstrate that as the aDM concentrates at the center of a halo, it redistributes the nearby CDM and baryons. Because of the lack of dark feedback, the aDM can cool, collapse, and deepen the halo potential at high redshift, thus shrinking the baryonic disk, skewing the stellar age distribution older, and reducing the half-light radii of satellite galaxies. These may point to potentially promising observables to constrain the aDM parameter space or delineate new discovery avenues. However, due to the computational cost of running Milky Way zoom-in simulations, these first works could not sufficiently scan over aDM parameter space, making it difficult to fully generalize the results. 

This paper presents the first cosmological hydrodynamical simulations to model the effects of aDM in isolated, classical dwarf galaxies with halo masses $\sim 10^{10}\msun$.  One primary advantage in working with dwarf-like systems is the ability to run a larger suite of simulations that vary over the aDM microphysics.  By sampling over this space, we identify a broad range of aDM parameters where the cooling is aggressive and leads to  universal properties of the halo distribution.  
In this region, aDM can significantly enhance the inner densities of dwarf galaxies: the aDM gas cools rapidly at high redshift and collapses into dark clumps, forming a cuspy inner region. Then, over the course of the halo's evolution, the inner region relaxes to a quasi-isothermal configuration. 
We develop simulation-informed fitting functions for the aDM density profiles, as well as analytical physical criteria for when these fitting functions apply. This allows our results to be generalized in aDM parameter space beyond the specific parameter choices we simulate.

This paper is structured as follows.  Section~\ref{sec:adm_intro} provides an intuitive overview of the aDM parameter space and introduces effective parameters that capture the relevant physics on galactic scales. Section~\ref{sec:sim_details} then reviews the simulation framework, summarizing the implementation of aDM. The key results of the aDM evolution in isolated dwarfs are summarized in Sec.~\ref{sec:sim_results}, with Sec.~\ref{sec:changing_parameters_results} expanding on the density evolution specifically. Section~\ref{sec:semi-analytics} presents simple fitting functions that can be used to describe the total dark matter distribution for $\sim 10^{10}\msun$ dwarfs. Section~\ref{sec:conclusion} concludes by discussing the observational implications of aggressively-cooling aDM in the centre of classical dwarfs. The Appendices provide further details on the aDM parameter space, the numerical robustness of the simulations, and tabulations of the morphology metrics and fitting function parameters detailed in Sections~\ref{sec:sim_results} and \ref{sec:semi-analytics}.

\begin{figure}[t!]
\includegraphics[
trim = {1cm, 1cm, 0cm, 0cm},
width=0.48\textwidth]{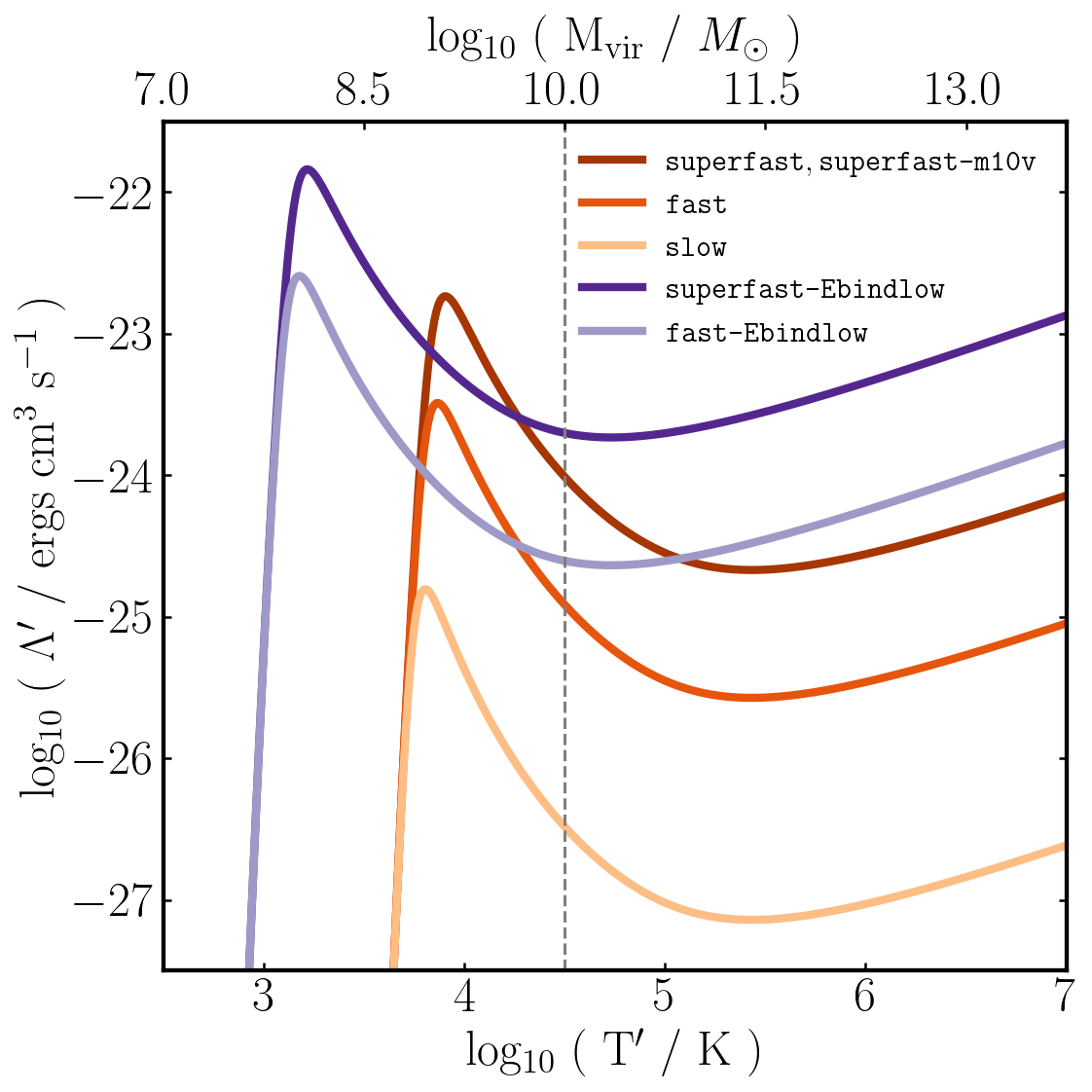}
\caption{
\label{fig:coolingcurves}
Plot of volumetric cooling rate $\Lambda'$ as a function of temperature for the zoom-in dwarf simulations  with aDM fractions $f'=6\%$. The exact parameters for each of these simulations are provided in Tab.~\ref{tab:ADMspecies}. We vary the overall dark cooling rate (the height of the curves) and the temperature at which the cooling rates become appreciable (referred to in the text as $\Tcutoff$). To compute the virial temperature for different halo masses, we use $\mpprime=\mproton$~\citep{Bryan1997}. All  parameters are $\order{1}$ ratios of the Standard Model values for reasons outlined in Section~\ref{sec:adm_aggressive_param_space} and Fig.~\ref{fig:coolingparamspace}. The vertical grey-dashed line shows the virial temperature for the halo mass $\sim 10^{10}\msun$. A greater height for the curve corresponds to more rapid aDM cooling while a lower $\Tcutoff$ allows dark atomic cooling to take effect at lower virial temperatures (thus at earlier times when the main halos are less massive). 
}
\end{figure}

\section{Mapping aDM Parameter Space}
\label{sec:adm_intro}


The minimal aDM model is a dark version of electromagnetism comprised of two oppositely-charged fermions: a dark electron $\eprime$ with mass $\meprime$ and a dark proton $\pprime$ with mass $\mpprime$. The fields are charged under a massless U(1) gauge field with a coupling strength described by the dark fine structure constant $\alphaprime$, which controls the dark electromagnetic field strength. There are two additional cosmological parameters needed to fully describe the model: $\emph{(i)}$ the mass fraction of aDM relative to the total dark matter density, $f'=\Omega_{\rm adm}/\Omega_{\rm dm}$, and $\emph{(ii)}$ the ratio of the dark to the standard CMB temperature, $\xiprime=T_{\rm cmb}'/T_{\rm cmb}$, which is constrained to be below 0.5 from CMB observations~\citep{Bansal:2022qbi}.  

The fermions $\eprime$ and $\pprime$ can form dark neutral Hydrogen, ${\rm HI'}$, 
with binding energy
\begin{equation}
\label{eq:Ebind}
    \Ebind = \frac{1}{2} \meprime \alphaprime^2 \,,
\end{equation}
as well as dark molecular Hydrogen, $\Htwoprime$. The dark Hydrogen gas cools via processes like collisional ionisation and excitation, recombination, and Bremsstrahlung.  The cooling rates, as calculated in~\cite{Rosenberg2017}, depend on the three microphysical parameters of the model and are functions of local number density, temperature, and ionization fraction. We also include the basic two-body molecular scattering processes $\Htwoprime-\eprime$, $\Htwoprime-\pprime$, $\Htwoprime-\Hneutralprime$ and $\Htwoprime-\Htwoprime$, using the rates and scalings of~\citet{Ryan:2021dis,Ryan:2021tgw}.

For illustration, Fig.~\ref{fig:coolingcurves} shows the volumetric cooling rate,
\begin{equation}    
\label{eq:Lambda}
 \Lambda' = \sum_i\,\langle{\delta E_i'\, \sigma_i' \, v'}\rangle\, n_{a_i}'\,n_{b_i}'/n'^2,
\end{equation}
where $i$ refers to the particular cooling process.\footnote{While Fig.~\ref{fig:coolingcurves} includes dark molecular processes, they are highly subdominant to the atomic processes. See Sec.~\ref{sec:sim_details} for more details.} The operator $\langle\ldots\rangle$ refers to a classical thermal average over a Maxwell-Boltzmann velocity distribution while $\delta E_i',\, \sigma'_i,\, v'$ are the energy losses, cross sections, and relative velocities of each process, respectively. The species involved in each process $i$ are given by the labels ${a_i}$ and ${b_i}$ (e.g. $a_i = e'$ and $b_i = p'$ for Bremsstrahlung) while the total number density $n'$ is defined as
\begin{equation}
\label{eq:nprime}
n'=n_{\rm HI'} + n_{\rm p'} + 2 n_{\rm H_2'} \, .
\end{equation}
 The number densities for each species are computed assuming $\mpprime \gg \meprime$, 
 collisional equilibrium between ionisation and recombination, the same temperature for dark protons and electrons, and uniform density of aDM within the halo's virial radius prior to the onset of collapse which is a suitable approximation for understanding how cooling depends on aDM parameters. For a more detailed discussion on how to compute $\Lambda'$ and the validity of these assumptions, see the appendices of \citet{Roy:2023zar} as well as \citet{Fan2013}.
 
 %
 %

 \begin{table*}[t!]\centering
\renewcommand{\arraystretch}{1.3}
\begin{tabular}{c|ccccccccccc}
  \Xhline{3\arrayrulewidth}
\textbf{Simulation} & \textbf{halo} & $\mathbf{\frac{{m_{\rm cdm}}}{{M_\odot}}}$ &  $\mathbf{\frac{m_{\rm adm}}{M_\odot}}$ & $\mathbf{\frac{m_{\rm b}}{M_\odot}}$ & $\boldsymbol{f'}$  & $\boldsymbol{\frac{\alpha'}{\alpha}}$ & $\mathbf{\frac{m_{p'}}{m_p}}$& $\mathbf{\frac{m_{e'}}{m_e}}$ &  $\mathbf{\frac{\tcool(z=0)}{\rm{14 \, Gyr}}}$ & $\log_{10}\betacool$ & $\mathbf{\frac{\Ebindeff}{\Ebindstd}}$ \\
\hline
$\Bhightlow$  & \mq  & $1200$ &  $650$ &  $250$  & 6\% & $0.58$ & 1 & 1.5  & 0.1 & $-3$& 0.5  \\
$\Bhighthigh$  & \mq & $1200$ & $650$ &  $250$  & 6\% & $0.41$ & 1 & 3   & 1 & $-2$ & 0.5\\
$\Bhightextreme$  & \mq & $1200$  &  $650$ &  $250$ & 6\% &$0.22$ & $1$ & $10$  & 10 & $-1$& 0.5 \\
$\Bhightlowmv$  & \mv  & $1200$ & $650$ &  $250$ & 6\% & $0.58$ & 1 & 1.5   & 0.1 & $-3$ & 0.5 \\
$\Bhighthighfhigh$  & \mq  & $900$ & $1300$ &  $250$ & 12\%  & $0.41$ & 1 & 3  & 1 & $-3$& 0.5  \\
$\Blowtlow$  & \mq  & $1200$ &  $650$ &  $250$ & 6\% & $0.45$ & $1$ & $0.5$  & 0.1 & $-5$ & 0.1 \\
$\Blowthigh$  & \mq & $1200$ &  $650$ &  $250$ & 6\% & $0.63$ & $1$ & $0.25$  & 1 & $-4$ & 0.1 \\
  \Xhline{3\arrayrulewidth}
\end{tabular}
\caption{\label{tab:ADMspecies} A summary of the dark matter properties for the simulations studied in this work. All simulations contain CDM, aDM, and baryons.  The total dark matter abundance (CDM + aDM) is $\Omega_\mathrm{dm} = 0.83 \, \Omega_{\rm m}$, where $\Omega_{\rm m}$ is the matter density.  The baryon abundance is $\Omega_{\rm b} = 0.17 \, \Omega_{\rm m}$ and particle mass is $m_{\rm b} = 250~M_\odot$.  The target zoom-in halos are $\sim 10^{10}~M_\odot$ field dwarfs (either \mq~or \mv), whose properties are detailed in the main text.  The CDM and aDM particle masses for the simulations, $m_{\rm cdm}$ and $m_{\rm adm}$ respectively, are listed in the table. 
For each simulation, the aDM constitutes a fraction $f' = \Omega_{\rm adm}/\Omega_{\rm dm}$ of the total dark matter abundance, with $f' = 6\%$ typically. Its cooling physics is determined by the dark proton mass~($m_{p'}$), the dark electron mass~($m_{e'}$), and the dark fine structure constant~($\alpha'$), which we calculate as $\alpha \cdot [(\Ebindeff/\Ebindstd) / (\meprime/\me)]^{1/2}$ to give the right effective binding energy. These quantities are listed relative to the corresponding Standard Model values~(unprimed).  We also report the 
instantaneous cooling time ratio $\tcool(z=0)/\thubble(z=0)$, as well as the effective parameters $\log_{10}\betacool$ and $\Ebindeff$, which are defined in Sec.~\ref{sec:adm_intro}. The first three simulations listed ($\Bhightlow$, $\Bhighthigh$, $\Bhightextreme$) are the main benchmarks and their names indicate the relative speed of aDM cooling.  The remaining four simulations explore variations on merger history, aDM fraction, and aDM binding energy on the final results. The value $\xiprime = T_{\rm cmb}'/\Tcmb$ varies from 0.1 to 0.2 between simulations; the results in the aggressively-dissipative parameter space ($\betacool \lesssim 10^{-2}$) are insensitive to these values, so they are not reported here (see App.~\ref{sec:varying_xiprime} for further details). 
}
\end{table*}

Table~\ref{tab:ADMspecies} lists the microphysical and cosmological parameters for the five aDM scenarios considered in this work and which are plotted in Fig.~\ref{fig:coolingcurves}.  All five cooling curves share the same characteristic shape.  At the lowest temperatures, the aDM gas is not ionized and there is no atomic cooling. As the temperature increases and eventually becomes comparable to the binding energy of the gas, the dark plasma is ionised and atomic cooling becomes efficient.
In this regime, the cooling rate is maximal when dark collisional excitation in the plasma dominates.  Beyond that point, the cooling rate diminishes until dark Bremsstrahlung takes over and causes $\Lambda'$ to monotonically rise. The temperature at which atomic cooling becomes efficient is proportional to the dark binding energy (Eq.~\ref{eq:Ebind}) while the overall cooling rate increases with increasing $\alphaprime$ and decreasing $\meprime$.

The general features of the aDM cooling curve dictate the phenomenology of the model on galactic scales. Specifically, the overall aDM cooling rate determines the efficiency of dark cooling on galactic timescales while the temperature at which atomic cooling becomes relevant~($\Tcutoff$) determines whether dark atomic cooling is relevant at the target halo masses and, if so, at what stage of the halo's evolution it begins.  

Understanding the galactic-scale phenomenology of aDM can therefore be facilitated by considering effective parameters that capture the qualitative features of the cooling curve. Compared to using all the microphysical and cosmological parameters, this approach can be highly advantageous as it both reduces the dimensionality of the problem and directly relates the aDM model space to observable quantities.  Here, we propose two effective parameters that adequately describe both the turn-on of aDM cooling and its overall strength across galactic timescales.

The first effective parameter captures whether the virial temperature of the halo, $T'_{\rm vir}$, is large enough for aDM cooling to be efficient.  This can be reframed as a question of whether the halo's virial mass, $M_{\rm vir}$, is large enough for cooling to turn on, given that  
\begin{equation}
\label{eq:Tvir}
    T'_{\rm vir}(z) = \frac{G M_{\rm vir}(z) \, \mpprime}{R_{\rm vir}(z)} 
\end{equation}
(assuming that $\mpprime \gg \meprime$), 
where $G$ is Newton's gravitational constant, $R_{\rm vir}$ is the virial radius of the halo, and $z$ is redshift.\footnote{$\mvir$ includes all particle species and refers to the total mass contained in the halo. In practice, we model $\mvir(z) \sim M_0 \exp(-\kappa z)$, where $M_0$ is the CDM halo mass at $z=0$ and $\kappa$ controls the rate of growth of the halo, as in \citet{Wechsler2002,McBride2009,Wang2009,Fakhouri2010,Genel2010,faucher-giguere2011,Benson2012,wu2013,Correa2014}). 
}  Thus, cooling turns on when $\Ebind/T'_{\rm vir} \lesssim 1$.  To this end, we define the effective dark binding energy for the neutral $\rm HI'$ gas as 
\begin{equation}
\label{eq:Ebindeff}
    \Ebindeff = \frac{E'_b}{m_{p'}/m_p} = \frac{1}{2}\frac{\meprime\alphaprime^2}{\mpprime/\mproton} \,.
\end{equation}
Dividing by the dark proton mass captures the scaling of $T'_\mathrm{vir}$ with $\mpprime$ for fixed $f'$, meaning that $\Ebindeff$ has the same parametric dependence on the three aDM microphysics parameters as $E'_\mathrm{b}/T'_\mathrm{vir}$, and the values of $\Ebindeff$ that allow for ionised dark atoms within a given halo mass remain the same regardless of $\mpprime$. The constant rescaling by $\mproton$ allows us to compare this effective parameter to the baryonic virial temperature.\footnote{In more detail, $\mproton \Tvir = \mpprime T_{\rm vir}$, so dividing $\Ebind$ by $(\mpprime/\mproton)$ naturally converts between $\Tvir$ and $T_{\rm vir}$.} Thus, if the baryonic virial temperature is significantly lower than $\Ebindeff$ (i.e. $k T_{\rm vir} \ll \Ebindeff$), then the aDM in the halo will remain neutral on average.

The second effective parameter captures whether aDM cooling, if it does turn on, is efficient enough to be relevant on galactic timescales.  The instantaneous dark cooling timescale of a halo, $\tcool$, is defined as
\begin{equation}
\label{eq:tcooleq}
    \tcool(\Tvir(z)) = \frac{\frac{3}{2}\,k \,\Tvir(z)}{\Lambda'(\Tvir(z)) \,n'} \, ,
\end{equation}
where $k$ is the Boltzmann constant.  For clarity, Eq.~\ref{eq:tcooleq} is written to explicitly show the dependencies on virial temperature and redshift.  Utilizing Eqs.~\ref{eq:Lambda} and~\ref{eq:Tvir} one can show the dependence of $\tcool$ on 
the aDM parameters:
\begin{equation}
\label{eq:tcoolscaling}
    \tcool \propto \frac{\Tvir}{\Lambda' \, n'} \propto \frac{\mpprime}{\Lambda' \, n'} \propto \frac{\mpprime^2}{\Lambda' \, f'} \, .
\end{equation}
As $f'$ decreases, $\tcool$ rises since the aDM number density falls and the dark plasma undergoes less frequent interactions. Increasing $\mpprime$ also increases $\tcool$, as this decreases the aDM number density and increases the average kinetic energy of each particle that has to be radiated away.

Importantly, $\tcool$ captures the cooling behavior at an instantaneous point in time.  However, as the universe expands and halos grow in physical (not co-moving) size, $\tcool$ at late times can underestimate the cooling rate, since it does not account for the possible collapse of the aDM halo at earlier times when average densities are greater. To include aDM cooling over a given halo's entire cosmological history, we define the following effective parameter:
\begin{equation}
\label{eq:betacool}
    \betacool = \left(\int_{\thubble(z=100)}^{\thubble(z=0)} d\thubble(z) \,\frac{1}{\tcool(\Tvir(z))}\right)^{-1} \, ,
\end{equation}
where $\thubble = 1/H(z)$ is the Hubble timescale and $H(z)$ is the Hubble constant.  The integral starts at $z=100$ since that is the starting point of the simulations and $\tcool(z\geq 100) \gg \thubble(z\geq 100)$.  The effective parameter $\betacool$ is an approximate quantity intended to be calculated to the correct order-of-magnitude. When $\betacool \ll 1$, then, on average, dark cooling is efficient over the relevant redshift range and mass evolution.  When $\betacool \gg 1$, then the dark cooling is inefficient over the halo's growth history.

\begin{figure*}[p!]
\includegraphics[
trim = {.5cm, 1cm, 1.5cm, 0cm},width=0.99\textwidth
]{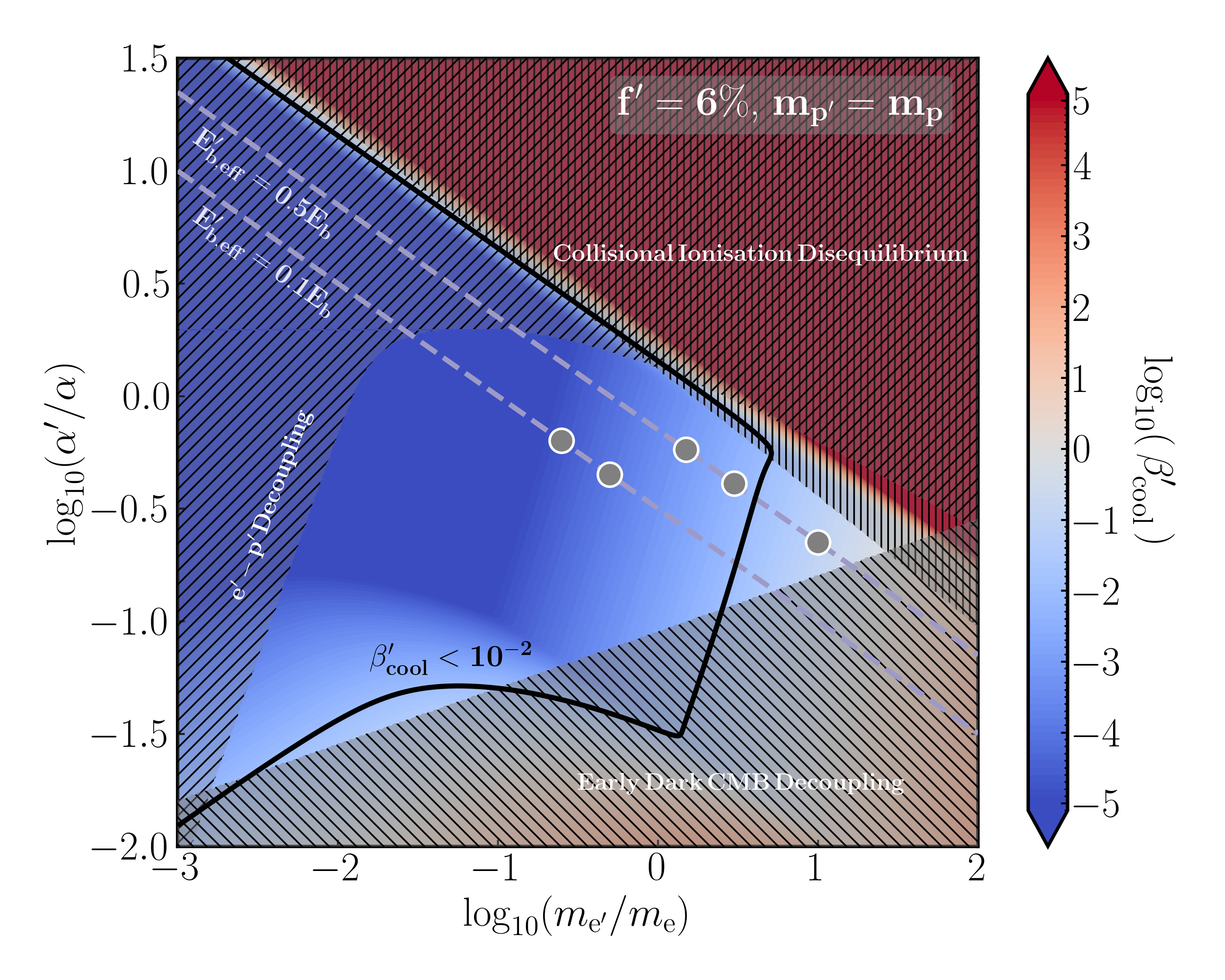}
\caption{
\label{fig:coolingparamspace}
A contour plot of the integrated cooling ratio $\betacool$ (Eq.~\ref{eq:betacool}) as a function of the dark electron mass, $\meprime$, and the dark fine structure constant, $\alphaprime$. Lower $\betacool$ values signify more efficient dark dissipation. The contour values are calculated using the fiducial simulation parameters: aDM mass fraction $f'=6\%$ and dark proton mass $\mpprime=\mproton$. In gray, we show the lines of constant $\Ebindeff/\Ebindstd = 0.5\,(0.1)$.  The dark black line denotes the boundary of the aggressively-dissipative parameter space, where $\betacool \sim 10^{-2}$. The hatched regions show where the simulation assumptions break down, as detailed in Sec.~\ref{sec:adm_intro}. The simulated aDM parameters, shown as grey dots, span the parameter values $\Ebindeff/\Ebindstd$ between 0.1 and 0.5 and $\betacool$ between $10^{-5}$ and 0.1. We do not probe lower dark binding energies as lower values result in later dark recombination, which may be in tension with cosmological observables~\citep{Bansal:2022qbi,Greene2024}, particularly for dark CMB temperatures $T_{\rm cmb}'/T_{\rm cmb} \geq 0.1$. Appendix~\ref{sec:adm_aggressive_param_space}  shows how the aggressively-dissipative contour alters with different $f'$ and $\mpprime$ values. This plot does not include molecular cooling, which is only relevant in the high $\betacool$ regions~\citep{Ghalsasi2017,Gurian:2022nbx} that are not simulated.
}
\end{figure*}


In sum, the two effective parameters $\betacool$ and $\Ebindeff$ capture the average aDM cooling efficiency as well as the temperature and mass scales at which dark atomic cooling becomes relevant. 
Therefore, the dimensionality of the aDM parameter space that must be considered for galactic processes is reduced to two parameters that describe cooling and $f'$ to capture the gravitational effects of the cooled aDM distribution on the baryonic and CDM components of the galaxy.\footnote{This is also physically reasonable considering that the dark proton mass should not matter explicitly as long as $\mpprime \gg \meprime$, as we assume. Further, the behaviour of aDM in the regime where it cools aggressively does not care about the value of  $\xiprime$.}


Figure~\ref{fig:coolingparamspace} plots $\betacool$ as a function of $\alphaprime$ and $\meprime$, assuming $f'=6\%$, $\mpprime = \mproton$, and the fiducial halo used in this work (described in Sec.~\ref{sec:sim_details}).\footnote{When calculating $\mvir(z)$ for this example, we take $\kappa=0.4$ and $M_0 = 10^{10}\msun$, as calibrated to the fiducial dwarf halo in $\Bhightlow$.}  We focus specifically on subdominant aDM scenarios whereby $f'\lesssim 10\%$, which is allowable given current CMB constraints \citep{Bansal:2022qbi} and theoretically motivated for various dark-sector models~\citep{Yuan2024,Ireland2024}.


The color scale in Fig.~\ref{fig:coolingparamspace} indicates the regions of parameter space where aDM cools very efficiently~(dark blue) to regions where it does not~(dark red).  The upper region of the plot is dark red because the binding energies are so high that no atomic cooling occurs; the halo mass never achieves a large enough $\Tvir$ to ionise the dark Hydrogen atoms. Along the direction of increasing $\meprime$ and decreasing $\alphaprime$, e.g., along curves of constant $\Ebindeff$, the dark cooling rate decreases and $\betacool$ increases. At very low $\Ebindeff$, $\betacool$ also increases, this time because dark atomic cooling becomes efficient at early times (high redshifts) where $\thubble$ is very low. So, the average virialised aDM in the halo quickly transitions to dark Bremsstrahlung 
and dark cooling becomes less efficient as the virial temperature of the halo grows since $\Lambda_{\rm brem.}'$ does not increase fast enough to radiate away the increased energy. 

The hatched areas in Fig.~\ref{fig:coolingparamspace} denote regions where the assumptions made in the simulation code break down. The upper region denoted ``Collisional Ionisation Disequilibrium" is where the aDM gas in the halo collisionally recombines and ionises too slowly for ionisation-recombination equilibrium to hold. The left region of the plot denoted ``$e'-p'$ Decoupling" is where the scattering rates between ionised dark protons and electrons are slower than the average halo cooling rates, so dark electrons and dark protons cannot share the same temperature. Finally, the lower region of the plot denoted ``Early Dark CMB Decoupling" is where $\alphaprime$ is so weak that the dark CMB likely decouples from the ionised aDM plasma before dark recombination at high redshifts---see \citet{Bansal:2022qbi} for more details. It is important to note that these regions are highlighted to show where our simulation assumptions break down and not because they are excluded by observational data. 

Five of the simulations presented in this work are shown by the gray points in Fig.~\ref{fig:coolingparamspace}, corresponding to $\Bhightlow$, $\Bhighthigh$, $\Bhightextreme$, $\Blowtlow$, and $\Blowthigh$.  These five cases are concentrated in the viable parameter space that can be simulated (non-hatched region of plot), with all aDM parameters roughly Standard Model-like. These simulations probe $\Ebindeff/\Ebindstd$ between 0.1 and 0.5, as well as $\betacool$ between 0.1 and $10^{-5}$. 

The solid black curve in Fig.~\ref{fig:coolingparamspace} outlines the regime inside of which $\betacool \leq 10^{-2}$.  As discussed in detail in Sec.~\ref{sec:sim_results} and  \ref{sec:changing_parameters_results}, the aDM gas for these dark halos cools so rapidly at high redshifts that their inner density profiles are almost identical. Nearly all the simulations presented in this work fall within this regime, except for $\Bhightextreme$, which provides an important contrasting case.

We do not simulate significantly lower $\Ebindeff$ in this investigation as it results in efficient cooling occurring at higher redshifts where the simulations are more poorly resolved.  Moreover, for the chosen $\xiprime$ values, lower $\Ebind$ results in later dark recombination, which may be in tension with the CMB and large-scale structure observations~\citep{Bansal:2022qbi,Greene2024}.  A more thorough investigation of cosmological constraints could elucidate how much of this parameter space is still viable, which would motivate running at higher resolutions to capture the relevant physics at early redshifts.  

For a discussion of how the aggressively-dissipative region varies with the other aDM parameters, see Appendix~\ref{sec:adm_aggressive_param_space}.

\section{Simulation Framework}
\label{sec:sim_details}

The suite presented in this work consists of seven cosmological hydrodynamical zoom-in simulations of isolated classical dwarf galaxies with mass $M_{\rm halo}\approx 10^{10}\msun$. Each simulation assumes a CDM plus aDM cosmology, varying over the effective aDM parameters $\betacool$ and $\Ebindeff$. The parameters for each simulation are provided in Tab.~\ref{tab:ADMspecies}. As a baseline, we take $\Ebindeff=0.5\Ebindstd$ and $f'=6\%$ and vary the dark cooling rate.
These simulations correspond to: $\Bhightlow$, $\Bhighthigh$, and $\Bhightextreme$, where the names indicate how (relatively) quickly the dark gas cools. The suite also includes simulations for a different host halo~($\Bhightlowmv$), aDM mass fraction ($\Bhighthighfhigh$), and dark binding energy ($\Blowtlow$ and $\Blowthigh$). For nearly all simulations, the CDM, aDM, and baryonic particle-mass resolutions are $1200\msun$, $650\msun$ and $250\msun$, respectively. The only exception is $\Bhighthighfhigh$, which has a CDM~(aDM) particle-mass resolution of $900\,~(1300)~\msun$ because the aDM (CDM) mass fraction is greater (lower) than that of the other simulations.

The main target halo(s) in the zoom-in region are picked from the standard Feedback In Realistic Environments~(FIRE) project, specifically the ``FIRE-2" suite, as described in~\cite{Wetzel2022,Hopkins2014}. For the fiducial halo, we use $\mq$, which has a total CDM mass at $z=0$ of $\sim 10^{10}\msun$ and is early-forming. To investigate the impact of halo-to-halo variance,  we also consider the $\mv$ halo, which has a mass of $\sim 10^{10}\msun$ and is late-forming. Both halos are derived from the initial conditions of the AGORA project~\citep{Kim2013,Kim2016} and are summarised in \citet{Hopkins2018}.

All simulations are conducted using the massively parallel, multi-physics code \texttt{GIZMO}~\citep{Hopkins2015}, which adopts the mesh-free Lagrangian Godunov ``MFM" method for hydrodynamics and an upgraded version of the Tree-PM solver for gravity~\citep{Springel2005}. CDM is simulated as collisionless particles with only gravitational interactions. The simulations use the FIRE-2 model for hydrodynamics and galaxy formation physics~\citep{Hopkins2014,Hopkins2017b,Hopkins2018}. The baryonic physics includes gas cooling down to the molecular regime~($\approx 10$--$10^{10}\K$)~\citep{Ferland1998-CLOUDY,Wiersma2009}, heating from a meta-galactic UV radiation background~\citep{Onorbe2016,FG2020} and stellar sources, star formation, as well as explicit models for stellar and supernovae feedback~\citep{Hopkins2014}. Baryonic gas particles are converted into collisionless star particles once the gas reaches densities exceeding $\sim 10^3\cm^{-3}$, contains non-zero molecular fractions~\citep{Krumholz2011}, and becomes Jeans-unstable and locally self-gravitating~\citep{Hopkins2013}. 


As first introduced in~\citet{Roy:2023zar}, aDM is implemented in \texttt{GIZMO} as a separate gas species that is decoupled from baryons except for gravitational interactions. The hydrodynamics of the aDM gas uses the same quasi-Lagrangian method as the baryonic gas does. The aDM atomic cooling functions are implemented according to~\citet{Rosenberg2017,Roy:2023zar} and assume that the dark CMB is the only dark radiation background. 

Once aDM gas particles become locally self-gravitating and Jeans-unstable, they convert into ``clump'' particles with the same mass as their parent gas particles over a free-fall time scale.\footnote{We additionally allow clump formation if the aDM gas particles have physical softening lengths $h^{\rm adm}_{\rm gas} \leq 1.5\pc$ and dark temperatures lower than $\Tcutoff$ as these particles are dense enough to hydrodynamically decouple.
This also prevents isolated and extremely dense dark molecular clouds from extending the computation time.} These dense clumps behave as collisionless particles at the resolved scale of the simulations. Unlike the baryonic gas, aDM contains no dark nuclear physics, so the aDM clumps do not undergo dark supernovae feedback. Physically, these clump particles are likely clusters of smaller and denser dark compact objects that are optically thick, such as dark stars but without any dark nuclear fuel---see~\citet{Gurian:2022nbx,Bramante2023,Fernandez2024} for further details on the dynamics and mass spectrum of dark compact objects. All aDM gas particles are optically thin with respect to the dominant atomic cooling rates. The only exceptions are extremely cold and dense aDM gas particles, 
 which may be opaque to dark photo-ionizing radiation from dark recombination. Because dark recombination is extremely subdominant compared to other cooling processes at low temperatures, we can safely neglect this effect.

 \begin{figure*}[t!]
\includegraphics[trim = {0cm, 0cm, 0cm, 0cm},width=0.99\textwidth]{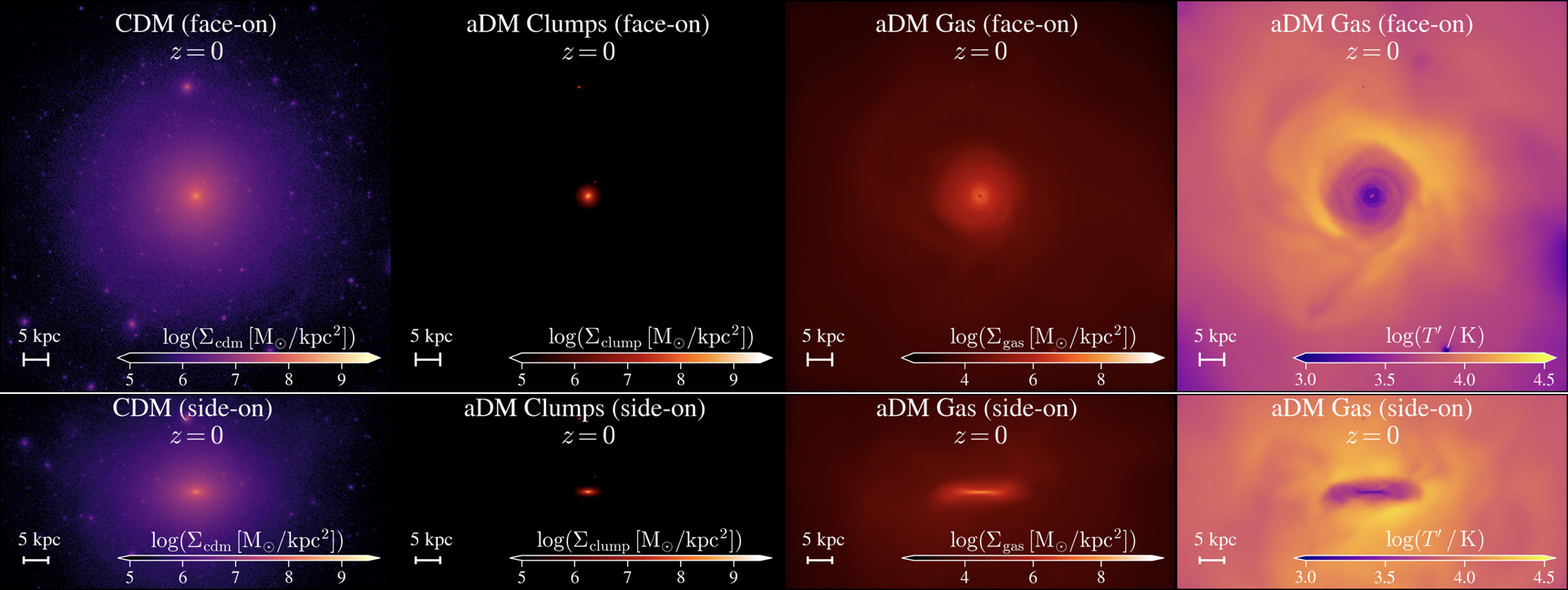}
\caption{
\label{fig:pretty_picture}
The first three columns (from the left) show the two-dimensional surface densities, $\Sigma$, of CDM, aDM clumps, and aDM gas for the $\Bhightlow$ simulation at redshift $z=0$. The present-day temperature projection of aDM gas is shown in the right-most column. The top row shows face-on projections while the bottom row shows side-on projections. CDM forms a halo-like distribution and dominates the overall dark matter density beyond $\sim 0.5\kpc$. The aDM clumps are centrally concentrated, dominating at smaller radii. The aDM gas forms a disk in the central halo, settling to a temperature of $T' \sim 0.5\times 10^4\K$, similar to the dark atomic cut-off temperature for this aDM model~(see Fig.~\ref{fig:coolingcurves}). The dark disk is surrounded by shock-heated aDM gas at the dark virial temperature of the halo.
}
\end{figure*}

This simulation suite expands the original aDM \texttt{GIZMO} implementation of~\citet{Roy:2023zar} to also include dark molecular cooling. We include the basic two-body molecular scattering processes $\Htwoprime-\eprime$, $\Htwoprime-\pprime$, $\Htwoprime-\Hneutralprime$, and $\Htwoprime-\Htwoprime$ using the rates and scalings of~\citet{Ryan:2021dis,Ryan:2021tgw}. This treatment conservatively uses a constant dark molecular fraction $x_{\Htwoprime} = n_{\Htwoprime}/n' = 10^{-6}$, where $n'$ is given in Eq.~\ref{eq:nprime}.  This fraction closely matches the dark primordial $\Htwoprime$ abundance at $z\sim 100$ for all the aDM parameters used in this study, as calculated with \texttt{DarkKROME}~\citep{Gurian:2021qhk,Ryan:2021tgw}. 
A more detailed dark molecular cooling implementation would not strongly affect the conclusions of this work, which focus on the collisionless dynamics of the aDM clumps post-collapse and assume no strong dynamics in the dark sector (e.g. no `dark stars' or `dark feedback').

The force softening for both the baryonic and aDM gas particles uses the fully-conservative adaptive algorithm from~\citet{Price2007}, where the gravitational force assumes the identical mass distribution as the hydrodynamic equations. 
The minimum physical gas softening length reached is $h_{\rm gas} = 0.3\pc$.  
The CDM force resolution of the simulations is set to $h_{\rm dm} = 1\pc$. Based on convergence tests summarized in App.~\ref{sec:sim_convergence}, we conservatively expect that predictions for the central host galaxy should be numerically robust at galactocentric distance scales of $r \gtrsim 0.06\kpc$. 

For the initial conditions, the transfer functions for baryons, CDM, and aDM are calculated using a modified version of \texttt{CLASS}~\citep{Blas2011,Bansal:2021dfh,Bansal:2022qbi}. \texttt{MUSIC}~\citep{Hahn2011} is then used to generate initial conditions at $z\sim 100$. We adopt the cosmological parameters ($h=0.702$, $\Omega_{\rm m} = 1 - \Omega_\Lambda = 0.272$, and $\Omega_{\rm b} = 0.0455$) to be consistent with the corresponding halos in the \texttt{FIRE} simulation suite \citep{Hopkins2018}. The other parameters that must be specified include the dark CMB temperature and the aDM mass fraction. We use different dark CMB temperature ratios $T_{\rm cmb}'/\Tcmb$ between $\sim$ 0.1 and 0.2, but show in App.~\ref{sec:varying_xiprime} that this variation has no significant impact on the properties of the central halo.

\section{aDM Evolution in Field Dwarfs}
\label{sec:sim_results}

The first three columns of Fig.~\ref{fig:pretty_picture} show the present-day projected density distributions of CDM, aDM clumps, and aDM gas for the $\Bhightlow$ simulation.  While CDM forms the expected halo-like distribution, the distribution of the aDM gas and clumps is far more distinctive.  The goal of this section is to review the morphology of these aDM components, focusing on how the aDM distribution evolves over the dwarf's history. Throughout, we reference several morphological parameters. The first is the flatness parameter, $\flatness = 1-c/a$, where $a$~($c$) is the semi-major~(semi-minor) axis of an ellipsoid.\footnote{To obtain $\flatness$, we iteratively compute the moment-of-inertia tensor of the aDM component in the central $10\kpc$ of the galaxy, obtaining its effective triaxial dimensions, until convergence is reached. Specifically, we repeat this process with particles within the derived ellipsoid boundaries until the boundary values converge to within $10\%$.} Note that $\flatness = 1$ is a thin-disk distribution and $\flatness = 0$ is a spherical distribution. The second parameter is $Z_i$, defined as the vertical height in which the $i^{\rm th}$ fraction of particles is contained in a cylindrical region of radius $R\leq 2 \kpc$.  Similarly, $R_{i}$ is the cylindrical radius in which the $i^{\rm th}$ fraction of particles is contained within a cylindrical region $|Z|\leq 2 Z_{9/10}$. These are the same morphology metrics used in \citet{Roy:2023zar}.

\begin{figure}[t!]
\includegraphics[trim = {5cm, .5cm, 1.5cm, 3cm},width=0.49\textwidth]{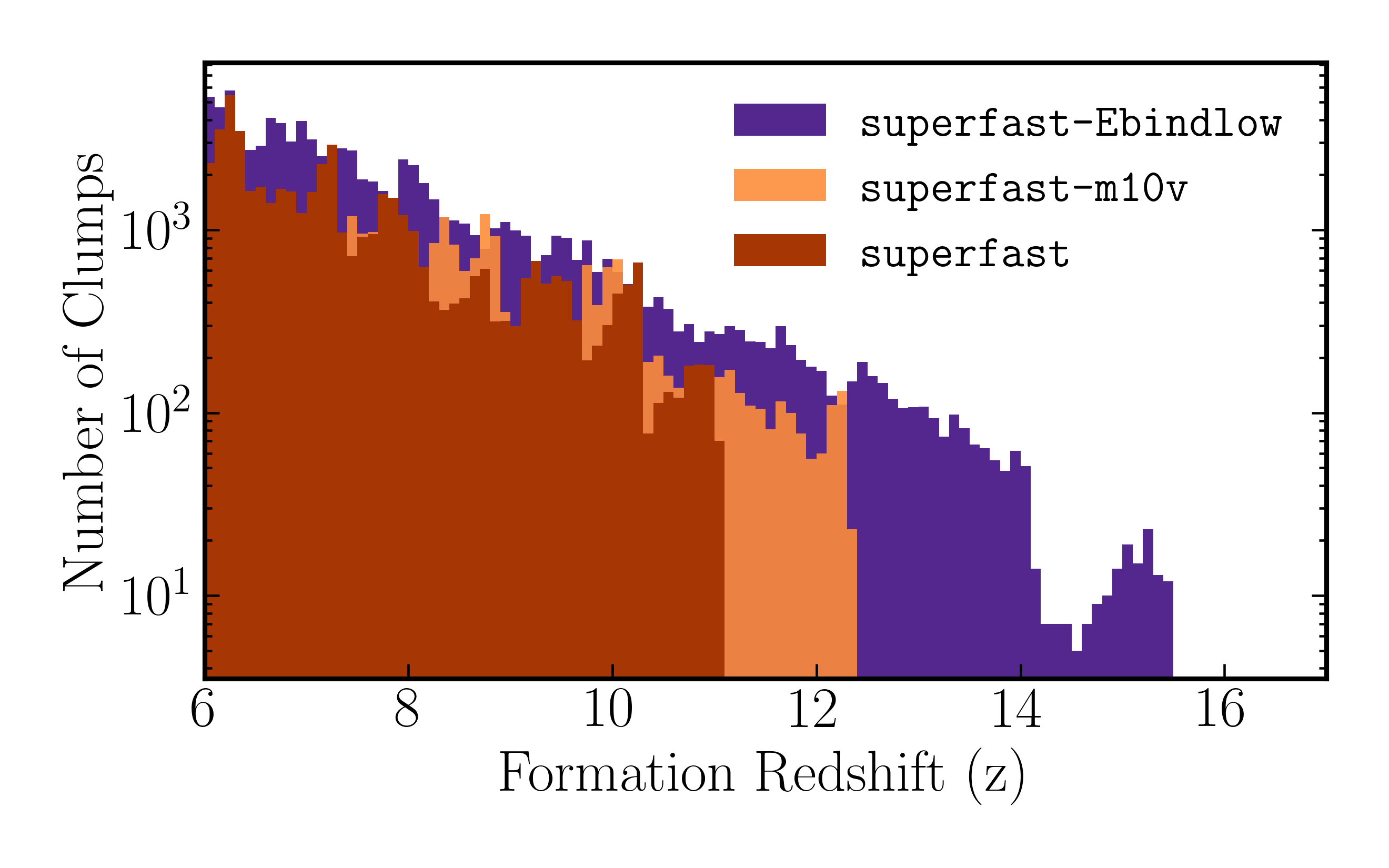}
\vspace{-1em}
\includegraphics[trim = {5cm, .5cm, 1.5cm, 3cm},width=0.49\textwidth]{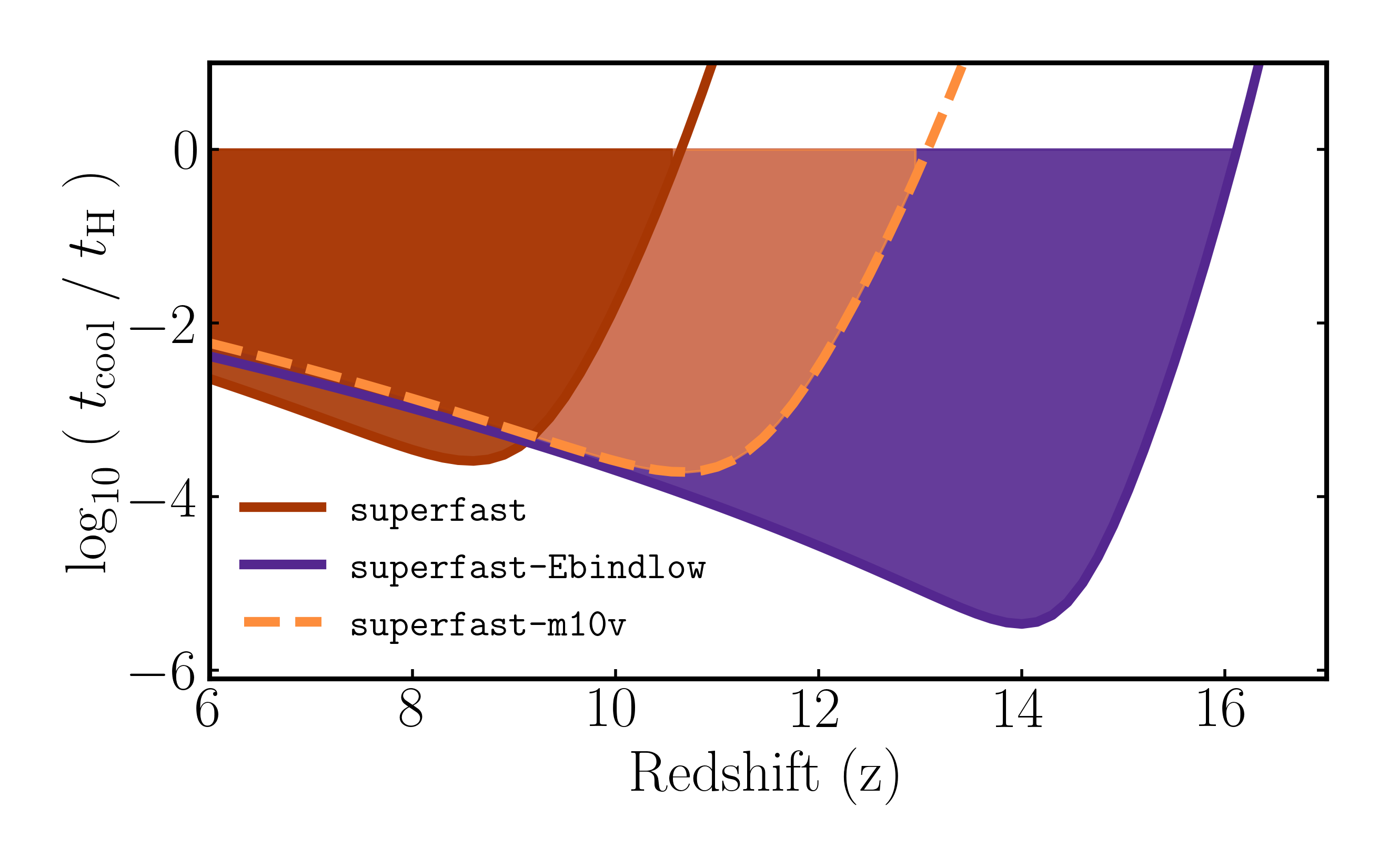}
\caption{
\label{fig:cooling_vs_clumpformation}
\emph{Upper panel}: The formation redshifts of all aDM clumps in the $\Bhightlow$ (maroon, opaque), $\Bhightlowmv$ (orange, translucent), and $\Blowtlow$ (purple) simulations. \emph{Lower panel}: The instantaneous cooling timescale, $\tcool/\thubble$, as a function of redshift for the most massive halo at each redshift. In all cases, clump formation rapidly turns on as the dark cooling becomes efficient. This corresponds to the redshift where the virial temperature of the dwarf halo is high enough to enable efficient atomic cooling (i.e. $\Tvir(z) > \Tcutoff$) and is thus dependent on both the dark binding energy as well as the halo's growth history.}
\end{figure}

The CDM distribution for the $\Bhightlow$ galaxy is spherically symmetric and dominates the total dark matter-projected surface density at large galactocentric distances. In contrast, the aDM clump distribution is centrally concentrated and elongated along the galactic plane, with a flatness parameter $\flatness=0.6$. The aDM clumps have a cylindrical half-mass containment radius of $\rhalf = 0.3\kpc$ and a half-mass containment height of $\zhalf = 0.08\kpc$. The aDM gas has a more distinctive disk-like distribution, with $\flatness =0.9$, $\rhalf=1.2\kpc$, and $\zhalf=0.04\kpc$. As shown in the right-most column of Fig.~\ref{fig:pretty_picture}, the gas disk temperatures are $\sim 5\times 10^3\K$, near the cutoff temperature in Fig.~\ref{fig:coolingcurves}, while the surrounding gas is shock-heated to $\sim 10^{4.5}\K$. The ratio of the enclosed aDM gas mass and aDM clump mass is 0.02 in the inner $0.5\kpc$, so aDM gas is strongly subdominant to aDM clumps in this region. 

Compared to the Milky Way-mass halos studied in \citet{Roy:2023zar}, the morphologies of the aDM components are similar. \citet{Roy:2023zar} found that aDM clumps dominate the inner densities of the host and form a thick-disk distribution with $\flatness\sim 0.5$, while the aDM gas forms a thin-disk distribution with $\flatness \sim 1$. There is also a similar relationship between aDM gas and clumps with the value of $\rhalf$ being greater for aDM gas but $\zhalf$ being greater for clumps. The main difference between the aDM distribution in Milky Way-mass halos and the distribution in these dwarf halos is that the values of $\rhalf$, $\zhalf$ are larger in the Milky Way-mass halos, as one would expect given the larger virial radii.  

\newcommand{\fgas}{f_{\rm gas}'}

The morphology metrics for the aDM clump and gas distributions for all simulations in the suite are provided in App.~\ref{sec:morphology_metrics}. 
The results are consistent with those of the $\Bhightlow$  simulation. Generally, the aDM clumps have a thick-disk-like distribution and dominate centrally. The aDM gas is distributed in a thin disk and is more extended radially than the clumps. The only exception is the $\Bhightextreme$ simulation in which the aDM gas disk has completely collapsed into clumps by $z=0$, likely because the surrounding gas doesn't cool fast enough to ``feed" the central disk. The fraction of aDM gas to clumps, $\fgas$, in the inner $0.5\kpc$ varies between 0.001 and 0.09, so the aDM clumps clearly dominate over gas. In the expanded region $r \leq 5\kpc$, $\fgas$ varies between 0.2 and 0.9, so the aDM gas distribution becomes more important.  However, the total aDM contribution is still sub-dominant to CDM in this region. 

Given that aDM clumps dominate the inner densities for all the halos in this suite, understanding their formation history is critical to understanding the main features of the aDM distribution in these dwarfs. 
The top panel of Fig.~\ref{fig:cooling_vs_clumpformation} shows the number of clumps that form at a given redshift, while the bottom panel shows the redshift dependence of the cooling timescale $\tcool/\thubble$.\footnote{To calculate $\tcool/\thubble$, we use the \texttt{Rockstar} halo finder~\citep{Behroozi2011} to obtain the virial mass of the most massive halo in the zoom-in region at each redshift and its virial temperature.} For all the simulations shown, clump formation begins roughly when dark cooling becomes efficient. This corresponds to the redshift where the most massive dwarf's halo mass is large enough for the virial temperature to ionise the dark gas. Thus, at high redshifts, dark atomic cooling becomes efficient and the dark gas can proceed to cool rapidly and collapse into clumps. As argued in \citet{Roy:2023zar}, all of these qualitative processes are analogous to what is seen in baryonic simulations with efficient cooling, but lacking a strong source of `feedback' from collapsed objects.

Figure~\ref{fig:cooling_vs_clumpformation} demonstrates that the clump formation depends on the choice of $\Ebindeff$, as expected. Comparing the total clump formation history and the cooling timescale evolution for the most massive halo in the simulation region at each redshift for $\Bhightlow$ and $\Blowtlow$, clump formation begins earlier when the effective binding energy is lower ($z\sim 15$ versus $z\sim 11$). For similar initial conditions, a lower $\Ebindeff$ means that the aDM gas can start cooling and forming clumps earlier in time when the virial mass (or, equivalently, virial temperature) of the most massive halo is lower. This process can be viewed through the lens of cooling curves~(see Fig.~\ref{fig:coolingcurves}) as aDM clump formation begins approximately when the virial temperature of the most massive halo in the simulated region becomes comparable to the cut-off temperature, $\Tcutoff$. Notably, a different halo growth history can also affect the onset of clump formation even if $\Ebindeff$ is unaltered. To demonstrate this, Fig.~\ref{fig:cooling_vs_clumpformation} also shows the result for $\Bhightlowmv$. In this case, the clumps form earlier than those in $\Bhightlow$. The most massive halo in $\Bhightlowmv$ reaches the critical virial temperature for efficient dark atomic cooling (i.e. $\Tvir \gtrsim \Tcutoff \approx 5\times 10^3\K$) sooner than that in $\Bhightlow$, and thus dark cooling and collapse begin earlier ($z\sim 12$ versus $z\sim 11$).

\begin{figure*}[t!]
\includegraphics[trim = {1.5cm, .5cm, 2.5cm, 1.5cm},width=0.99\textwidth]{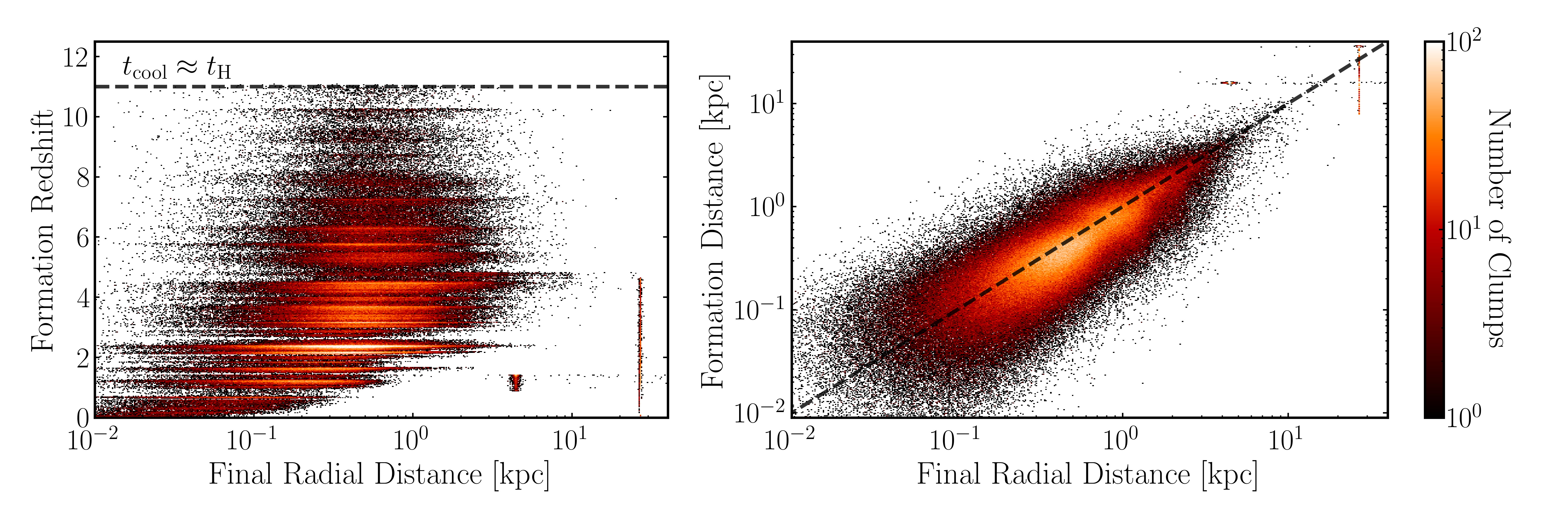}
\vspace{-1em}
\caption{
\label{fig:rform_zform}
\emph{Left panel}: Distribution of formation redshift and final distance from galactic center for aDM clumps in $\Bhightlow$. The dashed line $\tcool\approx \thubble$ represents the redshift value where the dark cooling timescale in the main halo first drops below the Hubble timescale. The formation redshift distribution of the central clumps shows that the  clumps tend to form in sharply-localised periods of redshift as opposed to smooth, gradual formation. The vertical streaks of clump formation show the presence of lower-mass subhalos which form dark clumps at different redshifts and formation distances to the host center, but are at a fixed final radial distance. \emph{Right panel}: Distribution of formation distance and final galactic radial distance for aDM clumps in $\Bhightlow$ along with a dashed, black 1:1 line. The majority of the clumps distributed in the inner $\sim 1\kpc$ remain at the same distance where they form with some scatter.
}
\end{figure*}

Once aDM clump formation begins, the process continues rapidly as there is no dark feedback mechanism to prevent runaway aDM gas collapse. Furthermore, the cooling timescales in Fig.~\ref{fig:cooling_vs_clumpformation} are significantly lower than $\thubble$ at high redshift with timescale ratios of $\lesssim 10^{-3}$, and even lower for $\Blowtlow$. The precise formation redshift distribution of aDM clumps within the inner $50\kpc$ of the main halo in $\Bhightlow$ is shown in the left panel of Fig.~\ref{fig:rform_zform}. The formation redshift distribution is bursty for clumps at galactocentric distances $\lesssim 10\kpc$, and periods of significant clump formation are often localised sharply in redshift. Since clump formation is a direct result of aDM gas collapse, this burstiness likely occurs because of mergers and rapid fluctuations in aDM gas accretion rates. 
The redshift at which clumps begin forming is $z\sim 11$ while the redshift at which $90\%$ of the final clumps have formed is $z\sim 1.3$. 

The right panel of Fig.~\ref{fig:rform_zform} shows the physical formation distance of the clumps alongside their final radial distances. The present-day location of clumps is strongly correlated with their formation distance, as indicated by the distribution's overall slope at final distances $r\leq 10\kpc$ closely following the black dotted 1:1 line, with a coefficient of determination ($R^2$) value of 0.64 between the logarithms of formation and final distances. This implies that most of the aDM clumps form \emph{in situ} and that the high central aDM densities we observe in Fig.~\ref{fig:pretty_picture} are already present around $z\sim 3$ when most of the clumps have already formed. 

The remaining simulations in the suite all have very similar clump evolution. Looking at the formation redshifts for clumps in the main halo at $z=0$, the earliest formation redshifts for the $\Bhighthigh$, $\Bhighthighfhigh$, $\Bhightlow$ and $\Bhightextreme$ simulations all occur in the range $z\sim 10$--$11$ because they share a consistent halo evolution history and the same $\Ebindeff$ as $\Bhightlow$. For a similar reason, the $\Blowthigh$ and $\Blowtlow$ simulations have approximately the same onset of clump formation, but at a higher redshift ($z \sim 16$) due to their lower value of $\Ebindeff$. In all simulations, $90\%$ of main halo clumps have formed by late redshifts in the range 0.1--1.3, with faster-cooling aDM simulations generally doing so before those with slower-cooling aDM (assuming the same $\Ebindeff$). 
The aDM in the main halo of $\Bhightlowmv$ is unique in that it evolves in a later-forming halo and the earliest clump formation redshifts are lower than $\Bhightlow$  ($z\sim 7$ instead of $z\sim 11$). The 90\% clump formation threshold in the main halo is also reached later ($z\sim 0.3$ instead of $z\sim 1.3$). Overall, all simulations show remarkably predictable and similar clump formation properties. The morphology metrics for all the simulations in the suite are provided in Tab.~\ref{tab:morphology_metrics_table} in App.~\ref{sec:morphology_metrics}.

\section{Dark Matter Density Profiles}
\label{sec:changing_parameters_results}

\begin{figure*}[t!]
\includegraphics[trim = {2.5cm, 1.5cm, 2.5cm, 1.5cm},width=0.49\textwidth]{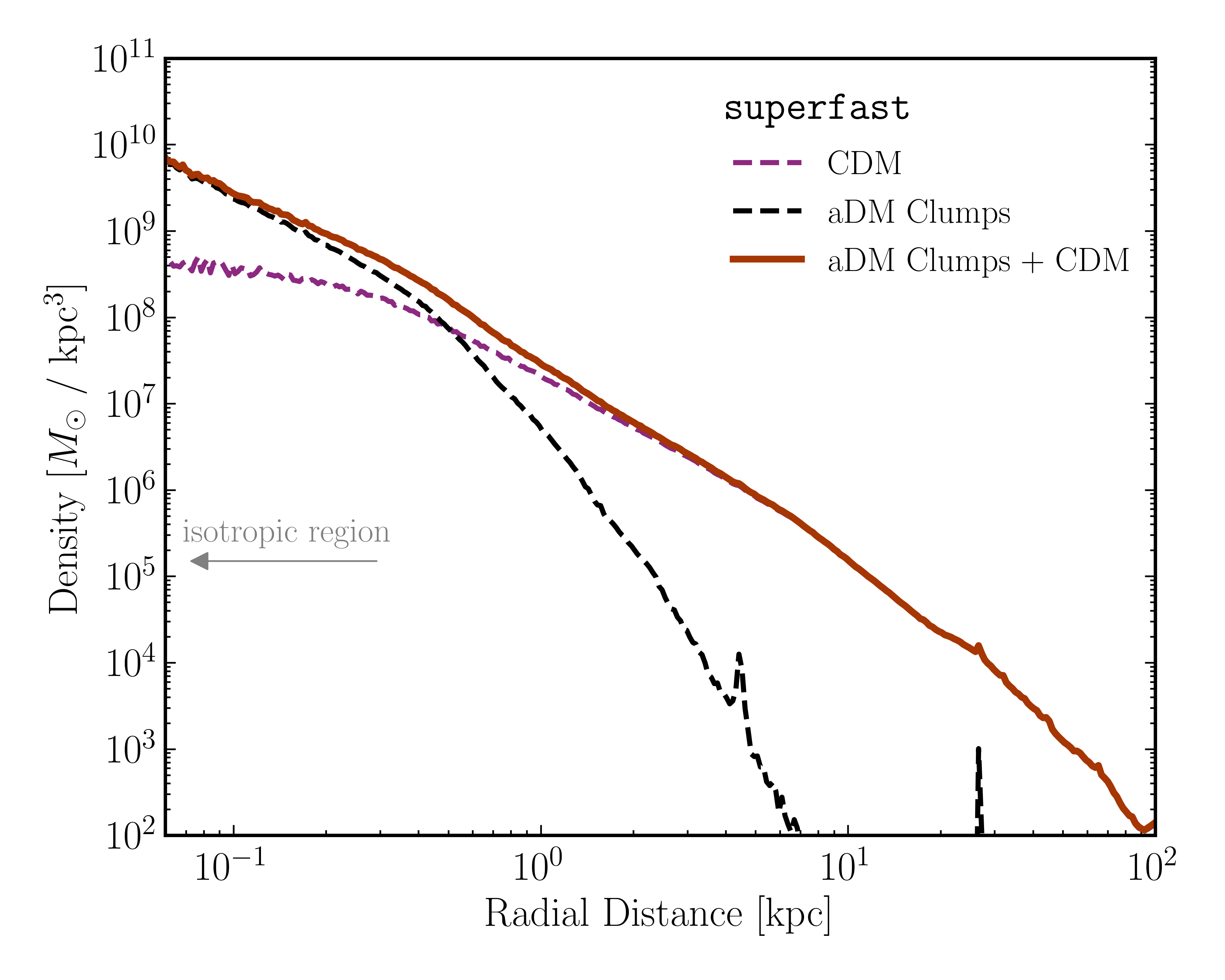}
\includegraphics[trim = {2.5cm, 1.5cm, 2.5cm, 1.5cm},width=0.49\textwidth]{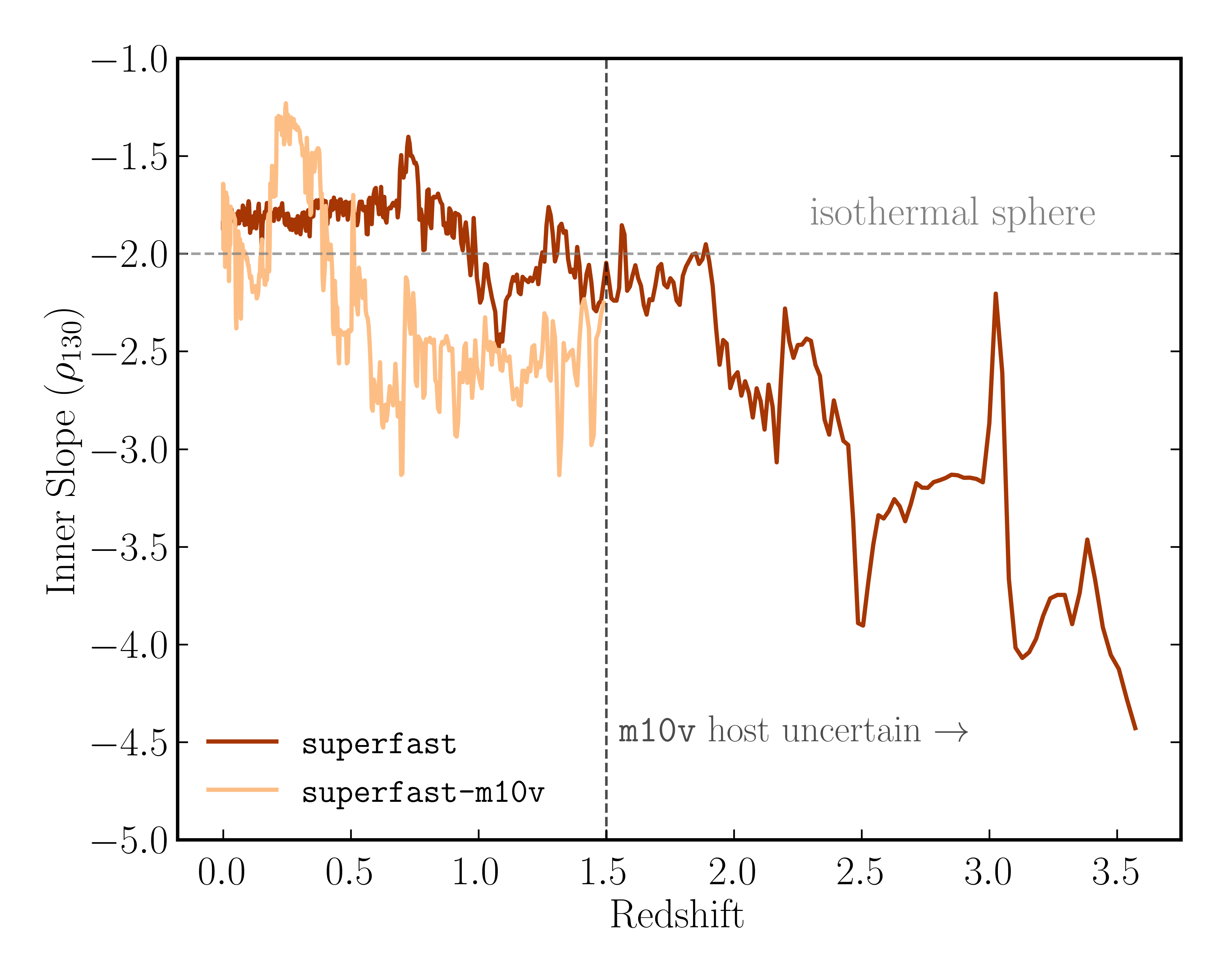}
\vspace{-1em}
\caption{
\label{fig:density_profiles}
\emph{Left panel}: Spherically-averaged present-day density profile of CDM~(purple) and aDM clumps~(black) for the $\Bhightlow$ simulation. The aDM clump densities dominate the inner regions of the halo where $r\lesssim 0.5\kpc$, while the CDM densities dominate the outer regions. We can ignore the aDM gas distribution because its density only begins to overcome that of clumps at distances $r\gtrsim 1\kpc$, where CDM is dominant. The total dark matter profile, shown in maroon, follows an almost constant $r^{-2}$ power law from $r\sim 0.06$--$10\kpc$, similar to the density profiles observed for weakly dissipative dark matter in \citet{Shen2021,oneil2023}. The same relation between aDM clumps and CDM holds for all other simulations with $f'=6\%$. \emph{Right panel}: Redshift evolution of the inner aDM density profile slope at $r=130\pc$ for the $\Bhightlow$ and $\Bhightlowmv$ simulations. For both cases, the inner slope transitions from extremely cuspy with inner slope $\lesssim -2.5$ to quasi-isothermal inner slopes of $\sim -2$. The inner slope exhibits significant noise and rapid changes which may be indicative of violent mergers and rapid fluctuations in gas accretion. By $z=0$, both simulations show an inner slope similar to that of a singular isothermal sphere.
}
\end{figure*}

Having quantified the basic properties of aDM evolution and morphology, this section focuses on the density profiles of the two dominant dark matter components: CDM and aDM clumps. The left panel of Fig.~\ref{fig:density_profiles} displays their present-day spherically-averaged density profiles (up to $50\kpc$) in the  $\Bhightlow$ simulation. The CDM density profile has an inner core that forms due to baryonic feedback~\citep[e.g.][]{Hopkins2018,Lazar2020}, and then the profile transitions to $\rhocdm \sim r^{-2}$ in the region $0.5\kpc \lesssim r \lesssim 10\kpc$.  Beyond that, $\rhocdm \sim r^{-3}$. 
The aDM clumps dominate over CDM within $r\lesssim 0.5\kpc$, where their spherically-averaged density profile follows a power law with $\rhoclump \sim r^{-2}$. 
The total density profile for both aDM clumps and CDM is shown in the left panel of Fig.~\ref{fig:density_profiles} by the solid maroon line. 
Notably, from the inner $0.06\kpc$ to $\sim 10\kpc$, this follows an almost constant $r^{-2}$ profile. 
Since the aDM clumps and CDM dominate the total density profile in this region, this nearly-constant power-law profile is a reasonable approximation of the total density profile, similar to the density profiles for dissipative dark matter in \citet{Shen2021,oneil2023}. 

 The right panel of Fig.~\ref{fig:density_profiles} shows how the inner slope of the spherically-averaged aDM clump profile for $\Bhightlow$ evolves with redshift. 
 The inner slope, which is measured at $r=130\pc$, softens from very cuspy values ($\sim -3.5$) to a final, quasi-isothermal value ($\sim -2$) through a series of rapid fluctuations in the density distribution from $z=3.5$ to the present day. 
This evolution may be a result of rapid aDM gas accretion or runaway aDM gas disk instabilities that may arise because there are no dark feedback sources to counteract gas disk collapse (see \citet{Toomre1964,Ostriker2011} for further theoretical discussion and \citet{Orr2017,Orr2020} for simulation-based investigations on stellar feedback and star formation). An additional effect may be violent relaxation caused by mergers, similar to the processes proposed to relax globular clusters and elliptical galaxies~\citep{King1966,bell1967,White1976,Binney2008,Zocchi2012,Baushev2014}.

\begin{figure*}[t!]
\includegraphics[trim = {2.5cm, 2.5cm, 7.5cm, 3.5cm},width=0.49\textwidth]{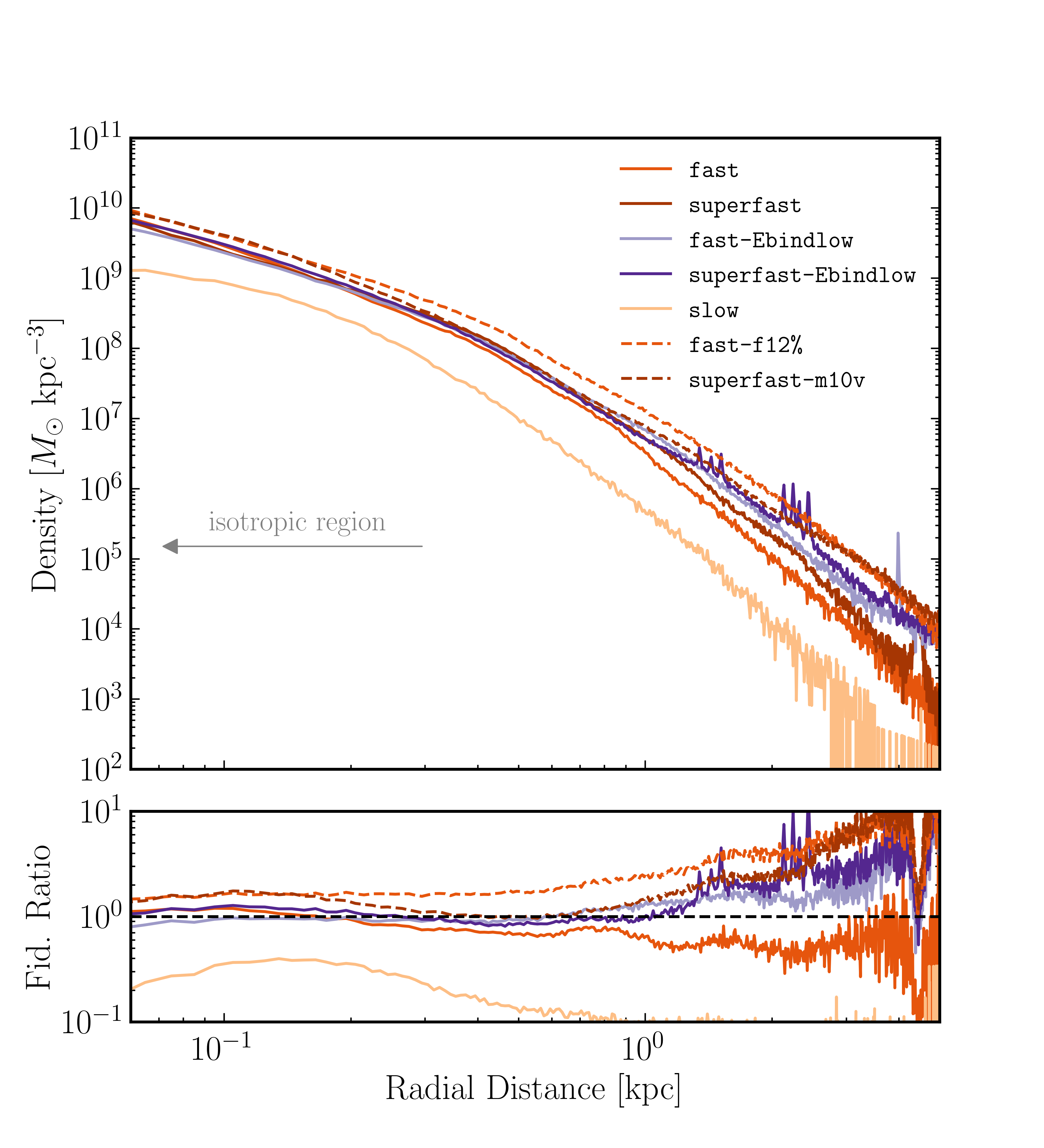}
\hspace{.1em}
\includegraphics[trim = {2.5cm, 2.5cm, 7.5cm, 3.5cm},width=0.49\textwidth]{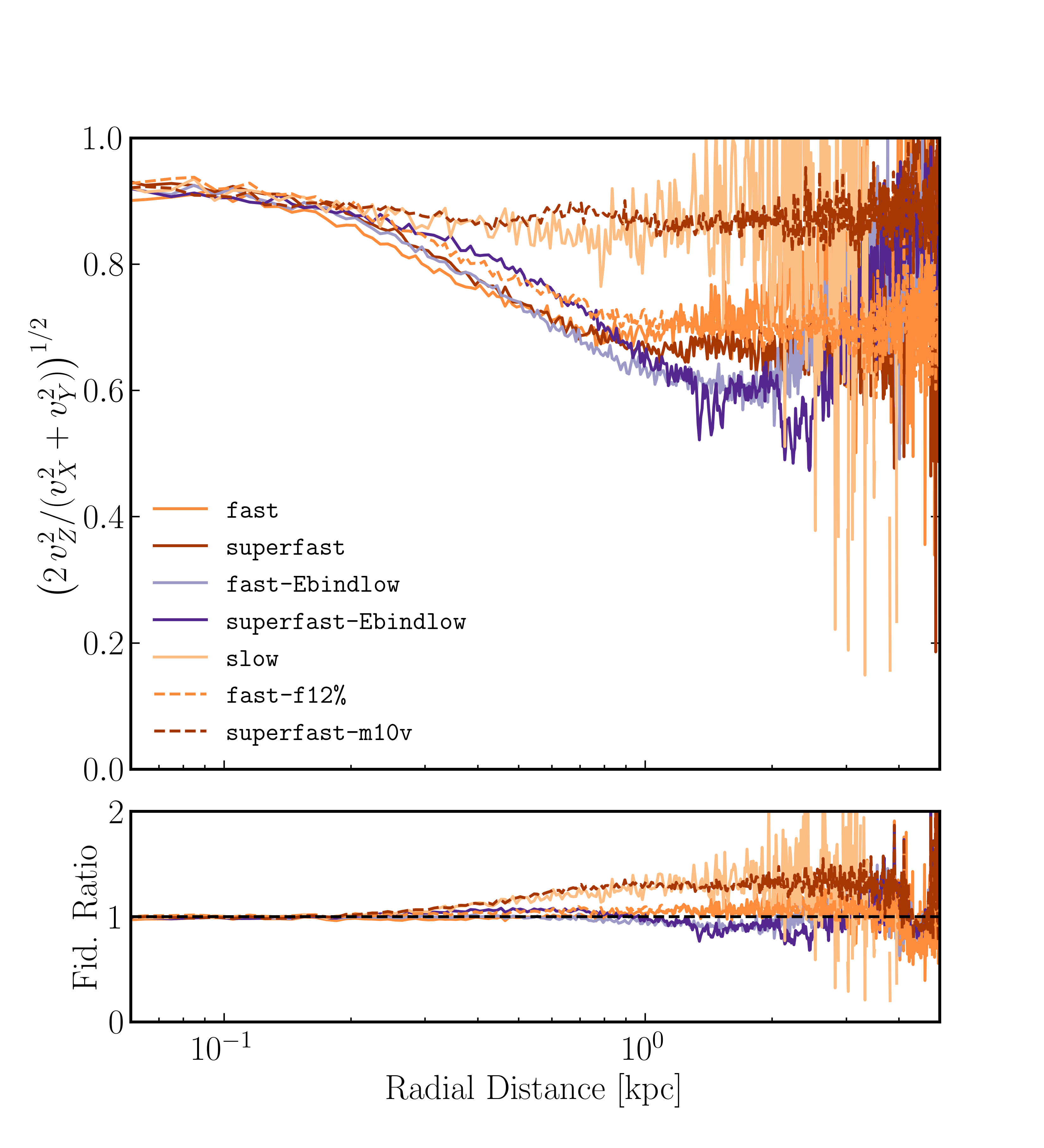}
\caption{
\label{fig:comparison_plots} \emph{Left panel}: The spherically-averaged aDM clump density profiles for all simulations in the suite. For simulations with $f'=6\%$ and with $\betacool \lesssim 10^{-2}$, the inner density profiles agree to a factor of $\lesssim 1.4$ (except for the inner region $r\lesssim 0.15\kpc$ for $\Bhightlowmv$ which has a fiducial ratio of $\sim 1.7$). For the $\Bhightextreme$ simulation, the inner densities are lower and more cored. For the $\Bhighthighfhigh$ simulation with $f'=12\%$, the density profiles are greater for all radii $r\lesssim 5\kpc$. Beyond galactocentric distances of $\sim 5\kpc$, the distributions are statistics-limited. \emph{Right panel}: Plot of the velocity isotropies of the aDM clump particles in each simulation, where ($v_X,\, v_Y$) refer to the cylindrical coordinates in the dark-disk plane whereas $v_Z$ refers to velocities in the vertical direction. For galactocentric distances $r\lesssim 0.3\kpc$, all density profiles are fairly isotropic. At distances greater than $\sim 0.3\kpc$, the anisotropies increase. The only exceptions are the $\Bhightextreme$ and $\Bhightlowmv$ simulations, for which the isotropies remain fairly constant up to $\sim 1\kpc$. The reason is that $\mv$'s evolution is different to that of $\mq$,  and so the final distribution of clumps is puffier, less elliptical, and more isotropic than that of $\mq$. For the $\Bhightextreme$ simulation, the aDM gas disk cools less efficiently and eventually fragments, so the gas disk's contribution to the inner clump densities is significantly inhibited. 
 The bottom panels provide the ratio of the relevant distributions with respect to the fiducial $\Bhightlow$ simulation.
}
\end{figure*}

Provided that $\betacool\lesssim 10^{-2}$, the inner aDM clump densities are fairly insensitive to nearly all the variations to the fiducial model explored in this work, as highlighted in the left panel of Fig.~\ref{fig:comparison_plots}. The simulations 
$\Bhighthigh$, $\Bhightextreme$, $\Blowtlow$ and $\Blowthigh$
capture variations in the two effective parameters of the model, varying them within $\betacool \in [10^{-5}, 10^{-1}]$ and $\Ebindeff/\Ebindstd \in [0.1, 0.5]$ (see Tab.~\ref{tab:ADMspecies}). The present-day density profiles of the aDM clumps are virtually identical for $r\lesssim 1\kpc$ across all variations in the cooling rates.  The only exceptions are the $\Bhightextreme$ simulation with $\betacool \sim 10^{-1}$, which has significantly lower inner densities and is more cored because the aDM cools less efficiently. Additionally, changing the fraction of aDM to $f'=12\%$ in $\Bhighthighfhigh$ has the predictable effect of enhancing the aDM clump densities at all radii by a factor of $\sim 2$ in the inner region~($r\lesssim 1\kpc$). 

Changing the halo growth history through a different choice of target ($\mv$ versus $\mq$) also has minimal impact on the final inner aDM distribution. In the left panel of Fig.~\ref{fig:comparison_plots}, the aDM clump densities for the $\Bhightlow$ and $\Bhightlowmv$ simulations agree to within a factor of $\lesssim 1.3$ over the entire region $0.2\kpc \lesssim r \lesssim 1\kpc$. The only major difference is within $r\lesssim 0.15\kpc$, where the $\Bhightlowmv$ clump densities are greater by a factor of $\sim 1.7$ likely because of differences in the halo's evolution history. The right panel of Fig.~\ref{fig:density_profiles} shows how the inner slope of $\Bhightlowmv$ evolves with redshift.  The $\mv$ host forms later than $\mq$, making it difficult to  reliably identify its main halo 
until $z \sim 1.5$.  However, the inner slope is still very cuspy at high redshifts ($\sim -2.5$ for $\Bhightlowmv$ at $z\sim 1.5$) and softens over time to a slope of $\sim -2$ by $z=0$, just like the clump distribution in the fiducial simulation. Thus, even though both halos have different evolution histories, the rapid cooling and collapse of aDM at high redshifts results in a collisionless, clump-dominated inner region that relaxes to almost the same quasi-isothermal profile by $z=0$.

Figures~\ref{fig:density_profiles} and \ref{fig:comparison_plots} show spherically-averaged density profiles, but---as already discussed in the previous section---the aDM clump distribution is not spherically symmetric at all radii. The right panel of Fig.~\ref{fig:comparison_plots} compares the ratio of cylindrical velocity components, $\sqrt{2v_Z^2/(v_Z^2 + v_Y^2)}$, where the $X$ and $Y$ directions span the dark-disk plane and $Z$ spans the vertical direction. 
In the inner $\sim0.3\kpc$, all simulations display a high ratio ($\gtrsim 0.8$) indicating that the aDM clump velocities are approximately isotropic. However, at greater galactocentric distances, the ratio diminishes as the ellipticity of the clump distribution becomes more apparent. The only exceptions are the  $\Bhightlowmv$ and $\Bhightextreme$ simulations, which are approximately isotropic at larger radii. In the case of $\Bhightlowmv$, the different evolution history results in a less-elliptical final distribution with a flatness parameter $\flatness$ which is $\sim 40\%$ lower than that of the fiducial simulation. For the $\Bhightextreme$ simulation, the lower cooling rate inhibits clump formation at late times when the aDM gas disk is more prominent and likely contributing to the ellipticity of the distribution (the final clump distribution is also approximately $\sim 20\%$ less flat than the fiducial simulation). However, all of the simulations investigated have isotropic clump distributions within the inner $\sim 0.3\kpc$ and since CDM dominates at galactocentric distances $r\gtrsim 0.5\kpc$, the combined CDM and aDM clump densities are spherically symmetric over the vast majority of distances in the region $0.06\kpc \lesssim r \lesssim 50\kpc$.

\section{Density Profile Fit Functions}
\label{sec:semi-analytics}

\begin{figure*}[t!]
\includegraphics[trim = {2.5cm, 2.5cm, 7.5cm, 3.5cm},width=0.5\textwidth]{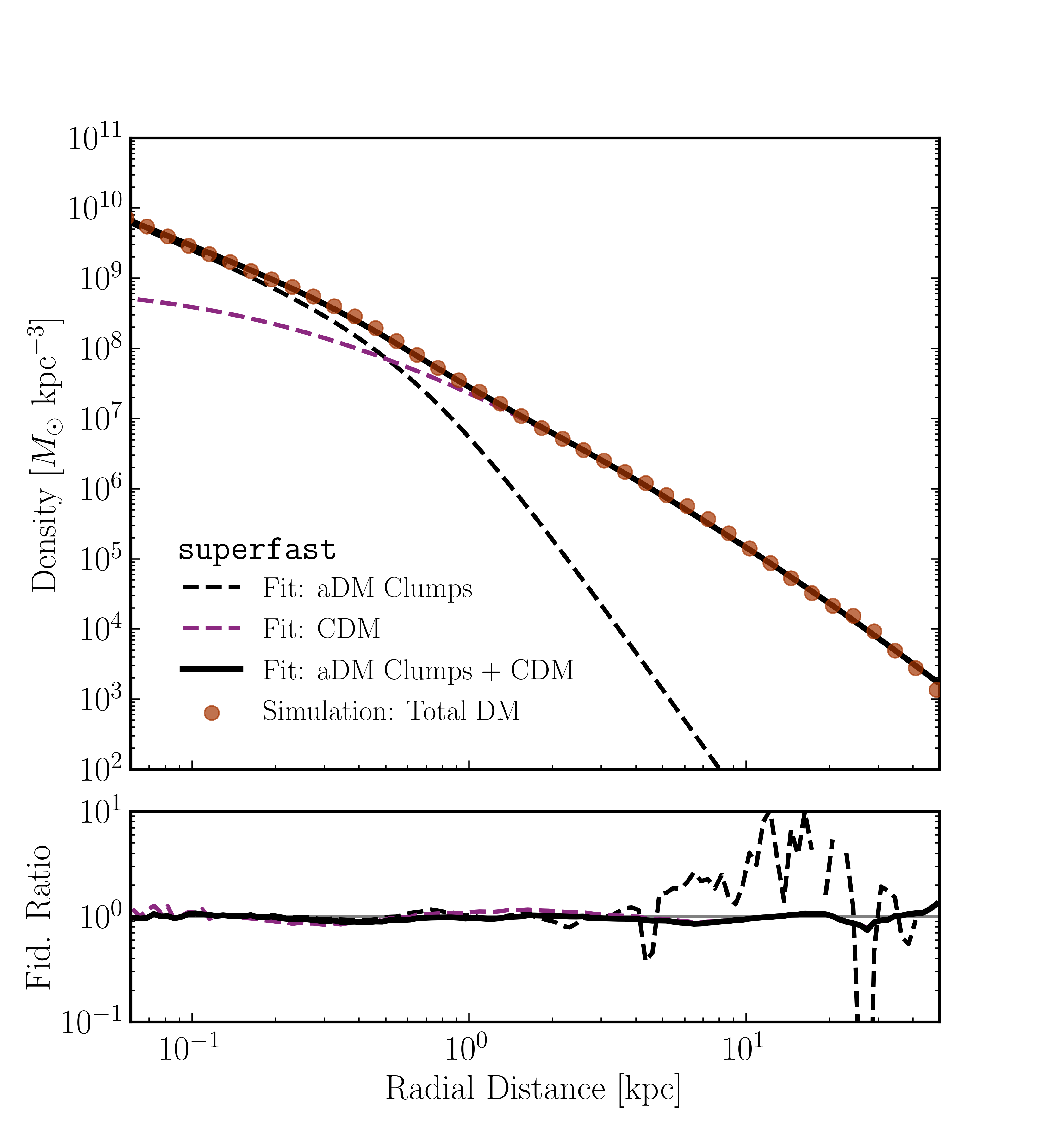}
\includegraphics[trim = {2.5cm, 2.5cm, 7.5cm, 3.5cm},width=0.5\textwidth]{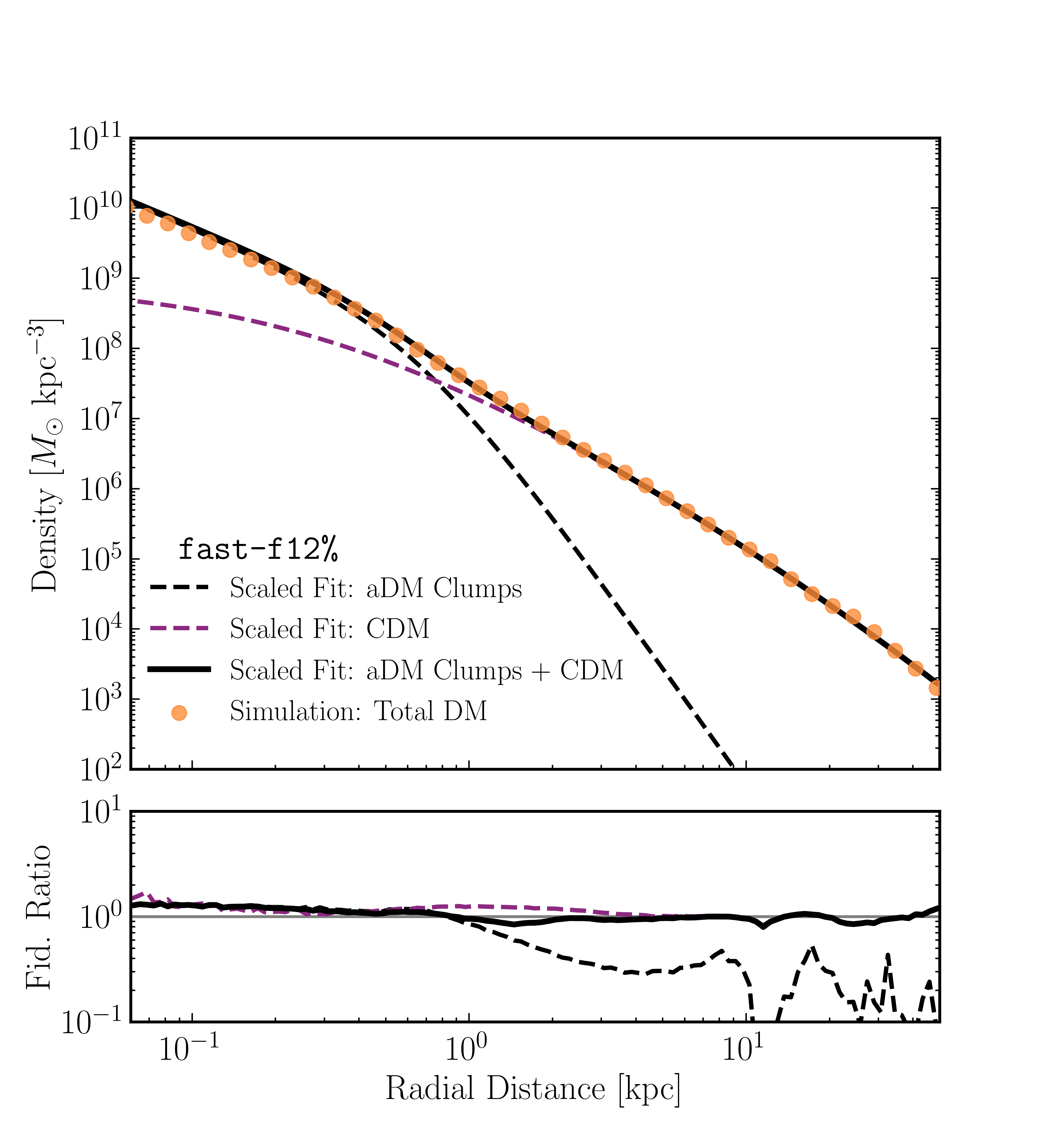}
\caption{
\label{fig:semi-analytic-fit} 
\emph{Left Panel:} The dashed lines show the aDM clump and CDM density-profile fitting functions for the fiducial simulation ($\Bhightlow$), while the solid black line is their total.  The maroon points correspond to the simulation density profile for all dark matter~(DM), including aDM clumps, aDM gas, and CDM.  The bottom panel shows the ratio of the fit densities to the densities in the simulation, labelled Fid.~Ratio. The overall fit agrees well with the simulation's total DM density profile, despite its simplifying assumptions (e.g. it neglects aDM gas). \emph{Right Panel:} Similar to the left panel, except that the points correspond to $\Bhighthighfhigh$.  Additionally, instead of fitting directly to the simulation, we use the fiducial fit from the left panel and scale the individual densities according to Eq.~\ref{eq:rhocdmscaling}. There is good agreement between the total fit and the simulation densities, with their ratio remaining reasonably low ($\lesssim 1.5$) over the entire radius range $0.06\kpc \leq r \leq 50\kpc$. For both $\Bhightlow$ and $\Bhighthighfhigh$, the aDM clump distributions are disk-like in the region $r\gtrsim 0.3\kpc$, so the spherically-symmetric aDM clump fit does not capture the actual geometry of the distribution. However, the aDM clump densities become subdominant to CDM densities at these distances, so it is possible to ignore these corrections to the fit model.
}
\end{figure*}

This section presents fitting functions that can be used to separately model the density distribution of CDM and aDM clumps in a 
$\sim 10^{10}\msun$ dwarf halo.  These functions apply specifically to the efficient dark cooling regime spanned by our simulation suite.  Motivated by this suite, the model assumes that the aDM clumps dominate in the inner-most region of the halo, forming a spherical, isotropic distribution.  At larger radii, the model assumes that CDM dominates over other components, enabling one to ignore the contributions of aDM clumps and gas, which are anisotropic in this regime.  Simple scaling relations enable one to extrapolate the fits to other aDM mass fractions easily. 

Following the approach of \citet{Lazar2020}, we model the CDM density profile using a cored Einasto profile
\begin{align}
\label{eq:rhocdm}
    \rho_{\rm cdm}(r) &= \rho_0 \, \exp\left(-\frac{2}{\alphae}\left[\left(\frac{r + r_c}{r_s}\right)^\alphae - 1\right]\right) \, ,
\end{align}
where $\rho_0$ is the density normalisation, $r_s$ is the scale radius, $r_c$ is the core radius, and $\alphae$ is the slope parameter. To model the aDM clumps, we use the generalised Navarro-Frenk-White~(NFW) profile~\citep{Navarro1997}, 
\begin{align}
\label{eq:rhoadm}
    \rhoadm(r) &= \rho_0' \left(\frac{r}{r_s'}\right)^{-\gamma} \left(1 + \left(\frac{r}{r_s'}\right)^2\right)^{-(\beta - \gamma)/2} \, ,
\end{align}
where $\rho_0'$ is the density normalisation, $\rsprime$ is the scale radius, $\gamma$ is the inner slope, and $\beta$ is the outer slope.  The ansatz of spherical symmetry should be sufficient in the inner regions of the galaxy ($r\lesssim 0.5\kpc$), as previously discussed.  Assuming that the aDM gas is sub-dominant and ignoring the anistropy of the aDM clumps at larger radii, this model assumes that the total dark matter density distribution is captured by
\begin{align}
\label{eq:rhototal}
    \rho_{\rm total}(r) &\approx \rho_{\rm cdm}(r) + \rhoadm(r) \, .
\end{align}


Because the total enclosed mass for a given species (CDM or aDM) must yield its correct cosmological abundance, the density normalizations can be related to the aDM mass fraction, $f'$, as follows: 
\begin{align}
    \label{eq:rhocdmscaling} 
    \rho_0 \propto (1-f')\quad \quad \text{and} \quad \quad \rho_0' \propto f' \, .
\end{align}
Therefore, for a given set of aDM parameters, one can easily obtain the total dark matter density profile for other values of $f'$ in the rapid aDM cooling regime.

\newcommand{\xa}{x_{\rm a}}

\newcommand{\rgas}{r_{\rm gas'}}

For each simulation in the suite, we separately fit the CDM and aDM clump distributions with Eqs.~\ref{eq:rhocdm} and~\ref{eq:rhoadm}. The best-fit aDM parameters for the fiducial simulation $\Bhightlow$ are provided in Tab.~\ref{tab:posteriors} while the fit parameters for all simulations (including CDM parameters) are provided in Tab.~\ref{tab:fitting_parameters_forall}.  The left panel of Fig.~\ref{fig:semi-analytic-fit} highlights the results for the $\Bhightlow$ simulation.  The dashed lines show the individual fits to CDM and aDM clumps, with the solid black corresponding to their sum.  The maroon points show the total dark matter profile (CDM, aDM clumps, and aDM gas) taken directly from the simulation.
The combined fitting function for aDM clumps and CDM match the total simulated dark matter densities to within a factor of $\lesssim 1.3$ over the range $0.06\kpc \lesssim r\lesssim 50\kpc$, despite ignoring the aDM gas contribution. The agreement is shown by the black line in the lower panel as a ratio of the fit density to the simulation's density. 

To verify the scaling relations in Eq.~\ref{eq:rhocdmscaling} with different $f'$ values (specifically $f'=12\%$), the right panel of Fig.~\ref{fig:semi-analytic-fit} plots the scaled fitting function from the fiducial simulation, $\Bhightlow$, and overlays the density profile from the $\Bhighthighfhigh$ simulation. The scaled fit captures the simulated density profile to within a factor of $\lesssim 1.4$ over the entire range, $0.06\kpc \lesssim r\lesssim 50\kpc$.

The fit uncertainties for the $\Bhightlow$ parameters are provided in Tab.~\ref{tab:posteriors}, but do not capture the systematic spread that arises across the full suite. As summarized in Tab.~\ref{tab:fitting_parameters_forall}, the aDM density normalisations $\rho_0'$ vary between $\sim 10^8$ and $10^9\msun\kpc^{-3}$ across all simulations in the suite with $f'=6\%$, and the scale radius $r_s'$ varies between $\sim 0.3$ and $0.7\kpc$. The aDM outer slope parameter $\beta$ varies between $\sim 4$ and $6$ and is thus quite susceptible to different aDM parameter choices. The inner slope parameter $\gamma$, however, is more constant across simulations, varying from the fiducial fit value of $1.7$ by $\sim 0.1$. The only exception is the simulation $\Bhightextreme$, where the aDM clumps are more cored and have a lower inner slope ($\gamma \sim 0.5$). 

\begin{table}[t!]
\footnotesize
\begin{center}
\renewcommand{\arraystretch}{1.5}
\vspace{0.5em}
\begin{tabular}{cccc}
  \Xhline{3\arrayrulewidth} 
 $\rho_0'\,[\msun\kpc^{-3}]$ & $\rsprime\,[\kpc]$ & ${\gamma}$ &  ${\beta}$\\
\cline{1-4}
 $(1.14\pm 0.13)\times 10^8$ & $0.65\pm 0.02$ & $1.67 \pm 0.03$ & $5.55\pm 0.06$ \\
  \Xhline{3\arrayrulewidth}
\end{tabular}
\end{center}
\caption{\label{tab:posteriors} Inferred aDM clump density profile fit parameters~(Eq.~\ref{eq:rhoadm}) based on the fiducial $\Bhightlow$ simulation. The fit parameters for the CDM profile and all the other simulations in the suite are provided in App.~\ref{sec:fitting_parameters_forall}. 
}
\end{table}

While this fitting model is advantageous for its simplicity, there are two key assumptions that are important to keep in mind when generalizing to different aDM parameters. Firstly, the fitting procedure only accurately describes aDM models in the parameter space where cooling is extremely efficient ($\betacool \lesssim 10^{-2}$). If the aDM parameters chosen do not cool this rapidly, then the fit parameters shown in Tab.~\ref{tab:posteriors} may not provide an accurate description of the final aDM distribution. Secondly, the fitting functions assume spherical symmetry, which may break down in the transition region where the aDM clump density is taken over by CDM, $0.3\kpc \lesssim r \lesssim 0.5\kpc$,  as demonstrated in Fig.~\ref{fig:density_profiles}. However, for lower $f'$ values, the CDM densities will begin to dominate over the aDM clump densities at lower radii, based on the scaling relations in Eq.~\ref{eq:rhocdmscaling}. Therefore, the assumption of overall isotropy will be valid over a larger range of radii for lower $f'$ values. 

\section{Discussion and Conclusion}
\label{sec:conclusion}

This work presents a new suite of cosmological hydrodynamical zoom-in simulations of $\sim 10^{10}\msun$ dwarf galaxies where the dark matter consists of both CDM and aDM. The aDM model is an example of a strongly-dissipative dark sector that forms a gas which can cool and collapse to form distinctive sub-galactic structures. Using the aDM module first developed for \texttt{GIZMO} by \cite{Roy:2023zar}, we simulated seven dwarf galaxies, spanning parameters such as the effective dark cooling rate and binding energy, as well as the aDM mass fraction.  We also considered variations on the target host to capture uncertainty on halo-to-halo variance.  This is the first such exploration of aDM parameters with hydrodynamical galaxy simulations, and it was made possible by the reduced computational cost of running dwarf zoom-ins relative to Milky Way zoom-ins. The first aDM simulations by~\cite{Roy:2023zar} focused on Milky Way-mass halos, and were thus limited to exploring only two aDM models.


By performing seven simulations that span a significant range of aDM parameters, and aided by analytical cooling arguments, we identify a distinctive region of aDM model space where the dark gas cools aggressively in $10^{10} \mathrm{M}_\odot$ classical dwarfs.  For these scenarios, the aDM becomes centrally-concentrated in the dwarf galaxy with a universal density distribution  that is insensitive to the specifics of the dark matter microphysics or the halo's evolution. This insensitivity arises because the aDM gas cools rapidly at high redshifts, collapsing into dense dark clumps that dominate the dwarf's central density. Subsequently, the evolution of these clumps is governed by collisionless gravitational dynamics, which results in a quasi-isothermal distribution.  In this central region, even a sub-dominant fraction of aDM, as low as 6\%, can significantly enhance the density of the dwarf galaxy to be approximately an order-of-magnitude greater than that in halos composed solely of CDM~\citep{Hopkins2018}.

For all simulations in the suite, nearly all the aDM gas ultimately ends up as clumps.  The fraction that does not is distributed as a thin dark disk at the center of the dwarf.  This dark disk comprises 0.1--9\% of the aDM within $0.5\kpc$ of the dwarf's center, but can reach 20--90\% of the aDM within $5\kpc$.  However, the dark gas contribution remains a sub-dominant fraction of the \emph{total} dark matter in this outer region, because CDM is dominant there.   These observations motivate a model for the dwarf dark matter density distribution, which includes a generalised NFW profile to capture the central distribution of aDM clumps and a cored Einasto profile to model the CDM. This model, despite its simplifying assumptions, reproduces well the total dark matter density profile obtained from simulations. 


To better understand the universal behavior of the aDM clump density in this aggressively-cooling regime, it would be beneficial to consider a greater number of target halos, ideally also varying over the baryonic physics implementation and aDM microphysics as well. This could potentially be achieved with an approach similar to \citet{camels2021,artemis2024,Rose2024}.  A larger sampling of dwarfs would allow one to understand the impact of halo-to-halo variance in setting properties of the aDM clump density distribution, including its scale radius and outer density profile. 

\begin{figure}[t!]
\includegraphics[trim = {2cm, 1.5cm, 0cm, 1cm},width=0.5\textwidth]{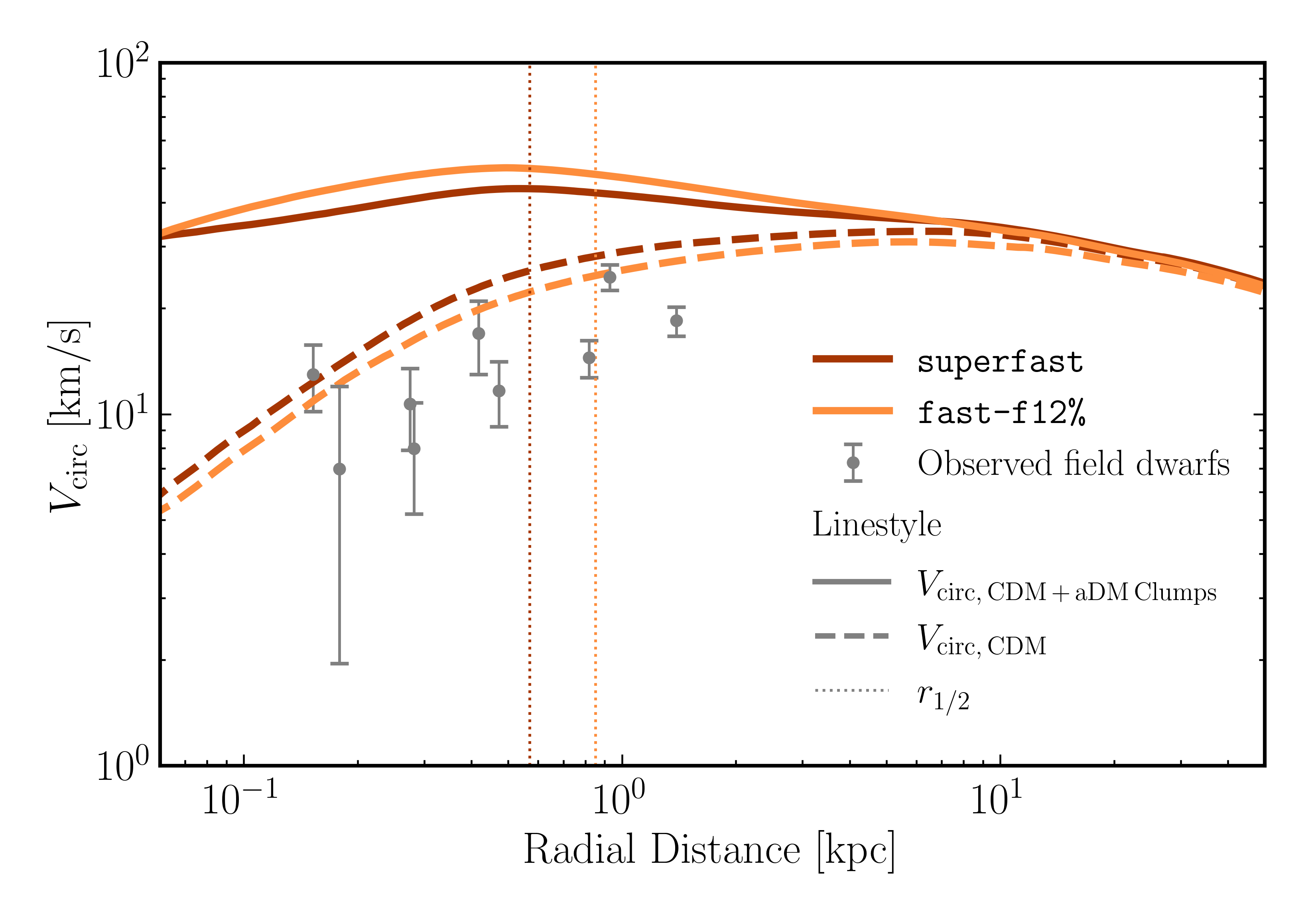}
\caption{
\label{fig:vcirc_profiles} 
Circular velocity profiles ($\sqrt{G M(<r)/r}$) for the $\Bhightlow$ and $\Bhighthighfhigh$ simulations, including those for CDM and aDM clumps and for just CDM. The total $V_{\rm circ}$ profiles are greater than the corresponding CDM profiles by a factor of $\gtrsim 3$ at radii $r\lesssim 0.1\kpc$ because the strongly dissipative aDM boosts the inner densities. We also display observed stellar half-light circular velocity data from observed local group field dwarfs \citep{Kirby2014,garrison-kimmel2014,Taibi2020}. While the CDM velocities are broadly consistent with the data, the velocity predictions from the aDM + CDM profile are greater by factors of $\gtrsim 2$ particularly at the simulation stellar half light values (displayed as dotted lines).
}
\end{figure}

The inner-density enhancements observed in the aggressively-cooling regime can potentially have extreme effects on field dwarfs, increasing baryonic star and gas rotation curves as well as the observed half-light circular velocities. As an illustration of these effects, Fig.~\ref{fig:vcirc_profiles} shows the circular velocity profiles for the $\Bhightlow$ and $\Bhighthighfhigh$ simulations ($\vcirc \equiv \sqrt{G M(<r)/r}$). Compared to the circular velocities computed using just the CDM enclosed masses~(dashed lines), the total circular velocities~(solid lines) are enhanced by factors of $\gtrsim 3$ for radii $r\lesssim 0.2\kpc$. The enhancement grows with larger aDM mass fractions.

Overlaid on the simulation curves in Fig.~\ref{fig:vcirc_profiles} are data for the circular velocities (inferred from line-of-sight stellar velocity dispersions) at stellar half-light radii for observed field\footnote{These dwarfs are at least $300\kpc$ away from the Milky Way or M31, following the definition in~\citet{garrison-kimmel2014}.} dwarfs in the Local Group. Specifically, we use the same compilation from~\citet{Shen2024}, including observations from \citet{Kirby2014} as well as updated values for the dwarfs with observed circular velocities from \citet{garrison-kimmel2014} and the Tucana dwarf \citep{Taibi2020}. Clearly, the observations prefer lower circular velocities than those in the two simulations shown. At the simulation stellar half-light radii $r_{1/2}$ in Fig.~\ref{fig:vcirc_profiles}, the simulation's circular velocities are greater than the observed values by a factor of $\gtrsim 2$. These preliminary results suggest that the aggressively-cooling aDM region may already be constrained by current observational data. Utilizing observed isolated dwarf galaxy rotation curves and stellar velocity dispersions may provide an avenue to thoroughly map out the constrained region and refine our understanding of aDM's role in shaping the structure of dwarf galaxies. This approach will also provide a benchmark for comparison against existing and upcoming observational data from the Spitzer Photometry and Accurate Rotation Curves~(SPARC) catalog~\citep{Lelli2016}, the Dark Energy Survey~(DES)~\citep{Abbott2005}, the Exploration of Local VolumE Satellites~(ELVES) survey~\citep{Carlsten2022}, the Satellites Around Galactic Analogs~(SAGA) survey~\citep{Geha2017}, the Nancy Grace Roman Space Telescope~\citep{Bailey2023}, the Large Synoptic Survey Telescope~(LSST)~\citep{LSSTCollaboration2009} at the Vera Rubin Observatory, the Analysis of Resolved Remnants of Accreted galaxies as a Key
Instrument for Halo Surveys~(ARRAKIHS) project~\citep{Guzman2022}, and the Merian survey~\citep{Luo2024}.


It is important to emphasize that the aDM parameter space is vast, and while this work has identified the aggressively-cooling parameter space as 
a distinctive target for systematic study and potential exclusion, much more remains to be explored.  Indeed, models with lower aDM mass fractions ($f'\lesssim 5\%$) and/or less efficient cooling ($\betacool > 10^{-2}$) lead to smaller central density enhancements and are thus much less constrained by dwarf velocity data.  This is demonstrated by the inner density profiles of the $\Bhightextreme$ simulation in our suite, whose aDM density profile is more cored and less centrally dense than the other simulations, suggesting much less conflict with present data. 


Additionally, there are regions of aDM parameter space that cannot yet be properly simulated or modeled. Investigating these regimes will require retooling our simulation code to better incorporate dark molecular cooling, dark plasma physics with decoupled dark protons and electrons, and  qualitatively different cosmological initial conditions. Moreover, the cooling physics for the parameter space where $\meprime \sim \mpprime$, far from the Standard Model-like region, has not yet been calculated. These regions of parameter space may have completely different galactic phenomenology, meriting their own dedicated studies. 

\section{acknowledgments}
The authors would like to acknowledge helpful conversations and feedback from Arpit Arora, Dylan Folsom, Zachary Gelles, Caleb Gemmell, Akshay Ghalsasi, James Gurian, Abdelaziz Hussein, Mariia Khelashvili, Lina Necib, Michael Ryan, Robyn Sanderson, Sarah Schon, Sarah Shandera, Maya Silverman, Oren Slone, and Linda Yuan. 

SR and ML are supported by the National Science Foundation~(NSF), under Award Number AST 2307789, and by the Department of Energy~(DOE), under Award Number DE-SC0007968. ML is also supported by the Simons Investigator in Physics Award. This research was supported in part by grant NSF PHY-2309135 to the Kavli Institute for Theoretical Physics~(KITP). NWM acknowledges the support of the Natural
Sciences and Engineering Research Council of Canada (NSERC; RGPIN-2023-04901). Support for PFH was provided by NSF Research Grants 20009234, 2108318, NASA grant 80NSSC18K0562, and a Simons Investigator Award. DC was supported in part by Discovery Grants from the Natural Sciences and Engineering Research Council of Canada and the Canada Research Chair program, as well as the Alfred P. Sloan Foundation, the Ontario Early Researcher Award, and the University of Toronto McLean Award. This work was performed in part at the Aspen Center for Physics, which is supported by National Science Foundation grant PHY-2210452. Numerical simulations were run on the supercomputer Frontera at the Texas Advanced Computing Center (TACC) under the allocations AST21010 and AST23013 supported by the NSF and TACC, and NASA HEC SMD-16-7592. This research is part of the Frontera computing project at the Texas Advanced Computing Center. Frontera is made possible by National Science Foundation award OAC-1818253.
Analysis of the simulations was performed on the Niagara supercomputer at the SciNet HPC Consortium. SciNet is funded by Innovation, Science and Economic Development Canada; the Digital Research Alliance of Canada; the Ontario Research Fund: Research Excellence; and the University of Toronto. The work reported on in this paper was also partly performed using the Princeton Research Computing resources at Princeton University which is consortium of groups led by the Princeton Institute for Computational Science and Engineering (PICSciE) and Office of Information Technology's Research Computing.

This research made extensive use of the publicly available codes
\texttt{IPython}~\citep{PER-GRA:2007}, 
\texttt{Jupyter}~\citep{Kluyver2016jupyter}, \texttt{matplotlib}~\citep{Hunter:2007}, 
\texttt{NumPy}~\citep{harris2020array}, \texttt{scikit-learn}~\citep{scikit-learn}, 
\texttt{SciPy}~\citep{2020SciPy-NMeth}, \texttt{SWIFTsimIO}~\citep{Borrow2020}, \texttt{unyt}~\citep{Goldbaum2018}, 
\texttt{gizmo-analysis}~\citep{Wetzel2020} and \texttt{Python Imaging Library}~\citep{clark2015pillow}

\bibliography{apssamp}

\newpage
\appendix

\setcounter{equation}{0}
\setcounter{figure}{0} 
\setcounter{table}{0}
\renewcommand{\theequation}{A\arabic{equation}}
\renewcommand{\thefigure}{A\arabic{figure}}
\renewcommand{\thetable}{A\arabic{table}}
\renewcommand*{\theHfigure}{\thefigure}
\renewcommand*{\theHtable}{\thetable}
\renewcommand*{\theHequation}{\theequation}

\section{Aggressively-Dissipative Parameter Space: Varying $\fprime$ and $\mpprime$}
\label{sec:adm_aggressive_param_space}

\begin{figure*}[h!]\centering
\includegraphics[trim = {3cm, 2.5cm, 1cm, 2.5cm},width=0.45\textwidth]{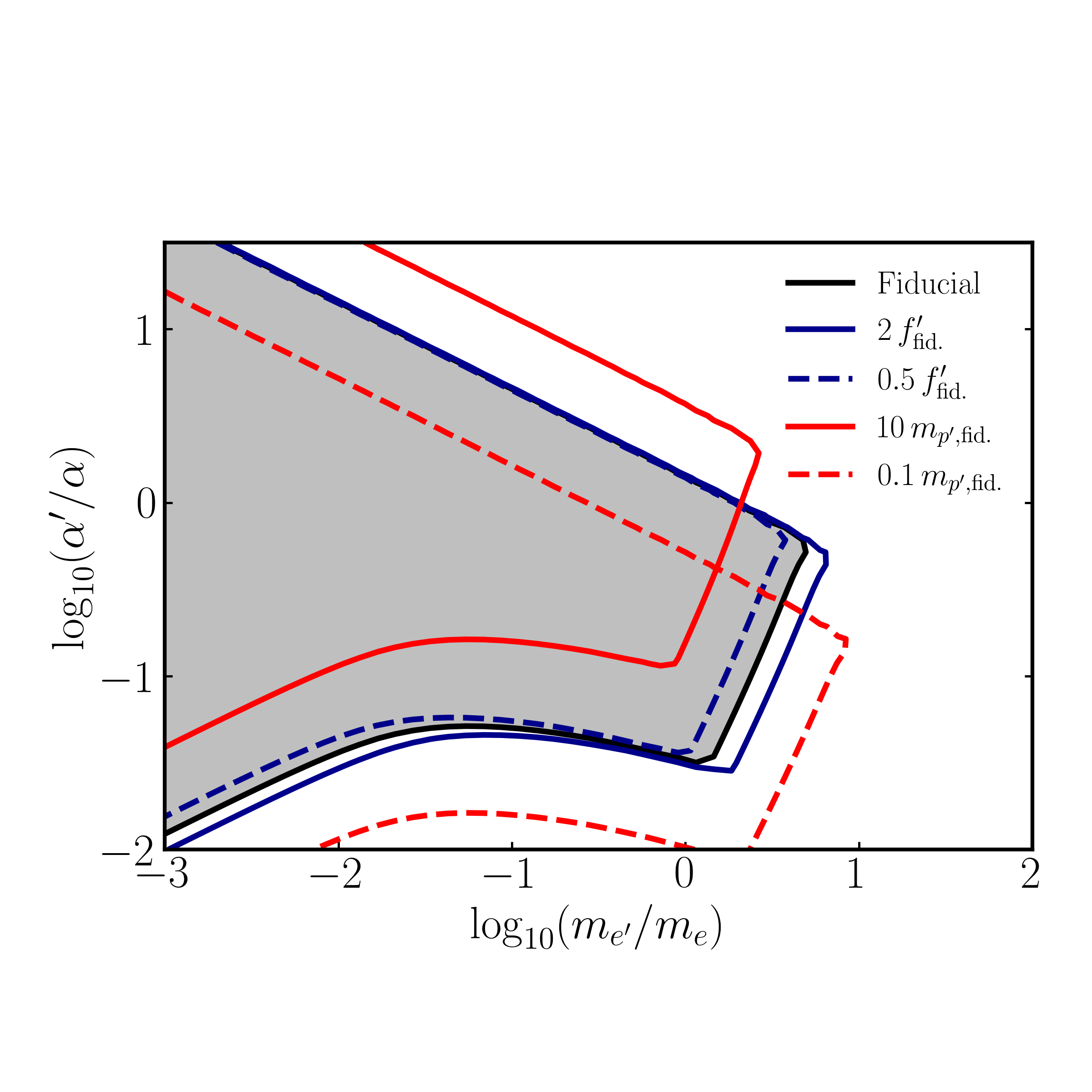}
\caption{
\label{fig:aggressive_paramspace_varyingfmp}
Contours of $\betacool = 10^{-2}$ plotted with changing $f'$ and $\mpprime$, as compared to the fiducial values ($f'=6\%$ and $\mpprime=\mproton$), labeled as $f'_{\rm fid.}$ and $m_{p',\rm{fid.}}$. The solid~(dashed) blue lines correspond to $2~(0.5)~f'_{\rm fid.}$, while the solid~(dashed) red lines correspond to $10~(0.1)~m_{p',\rm{fid.}}$. 
}
\end{figure*}

Figure~\ref{fig:coolingparamspace} shows the cooling parameter space for aDM for the fiducial simulation parameters $f'=6\%$ and $\mpprime=\mproton$. The fiducial boundary of the aggressively-dissipative region ($\betacool \leq 10^{-2}$) is outlined in black. Fig.~\ref{fig:aggressive_paramspace_varyingfmp} shows how this boundary changes with $f’$ and $\mpprime$. 

The upper edge of the contour delineates the maximum $\Ebind$ value above which the fiducial halo cannot ionise the aDM gas.  The only way to access aDM parameters with these large binding energies is to increase the virial temperature of the halo $\Tvir$, which can be done by increasing $\mpprime$.  For this reason, the solid red contour in Fig.~\ref{fig:aggressive_paramspace_varyingfmp}, which corresponds to a tenfold increase in the fiducial $\mpprime$, shifts upwards relative to the black contour.  Changing the aDM mass fraction has no effect here, because it does not change $\Tvir$ and no increase in $f'$ can compensate for the low cooling rate $\Lambda'$ when $\Tcutoff > \Tvir$. 

Decreasing $\Ebind$ has the effect of shifting the peak of the atomic cooling curve to lower $\Tvir$, leading to faster cooling at larger redshifts (i.e. at earlier times when $\mvir$ is lower). Since $\betacool$ can be thought of as the inverse of $d\thubble/\tcool$ summed over time, shifting the minimum $\tcool$ to greater redshifts results in lower $d\thubble/\tcool$ values and thus a greater $\betacool$. In this regime, $\betacool$ will only decrease by decreasing the instantaneous cooling time---either by increasing $f'$ or decreasing $\mpprime$. For this reason, the lower edges of the solid blue contour and the dashed red contour are shifted downwards towards lower $\Ebind$ compared to the black contour.

Keeping $\Ebind$ constant and moving to larger $\meprime$ and lower $\alphaprime$ decreases the overall cooling rate (i.e. the height of the cooling curve).  The only way to compensate for this is to increase the aDM number density, which can be achieved by either increasing $f'$ or decreasing $\mpprime$. As a result, the rightmost edges of the dashed red line and the solid blue line, corresponding to a tenfold decrease in $\mpprime$ and a doubled $f'$ respectively, are shifted towards greater $\meprime$ and lower $\alphaprime$ relative to the fiducial black contour.

\vspace{0.35in}

\setcounter{equation}{0}
\setcounter{figure}{0} 
\setcounter{table}{0}
\renewcommand{\theequation}{B\arabic{equation}}
\renewcommand{\thefigure}{B\arabic{figure}}
\renewcommand{\thetable}{B\arabic{table}}
\renewcommand*{\theHfigure}{\thefigure}
\renewcommand*{\theHtable}{\thetable}
\renewcommand*{\theHequation}{\theequation}

\newpage

\section{Impact of Dark Photon Temperature}
\label{sec:varying_xiprime}

This section explores the impact of changing the ratio of the dark Cosmic Microwave Background~(CMB) temperature to its standard value, $\xiprime = T_{\rm cmb}'/\Tcmb$. This is a necessary input for generating the initial conditions, as it effectively controls the suppression of the initial power spectrum due to dark acoustic oscillations~\citep{Bansal:2021dfh,Bansal:2022qbi,Cyr-Racine:2013fsa}. To test the effects of different levels of suppression, we vary the dark CMB temperature ratio for two simulations: $\Bhighthigh$ and $\Blowthigh$. The reason for this choice is that these two simulations are the slowest cooling for the two values of $\Ebindeff$ considered within the region $\betacool \lesssim 10^{-2}$. Thus, if varying $\xiprime$ does not result in appreciable differences in the final density profiles of these simulations, the same should be true for the more rapidly-cooling simulations in the aggressively-cooling regime. Further simulations would be needed to confirm the $\xiprime$ dependence outside the aggressively-cooling regime, including for the $\Bhightextreme$ simulation. 

\begin{figure*}[h!]
\includegraphics[trim = {1.5cm, 2cm, 1.5cm, .5cm},width=0.49\textwidth]{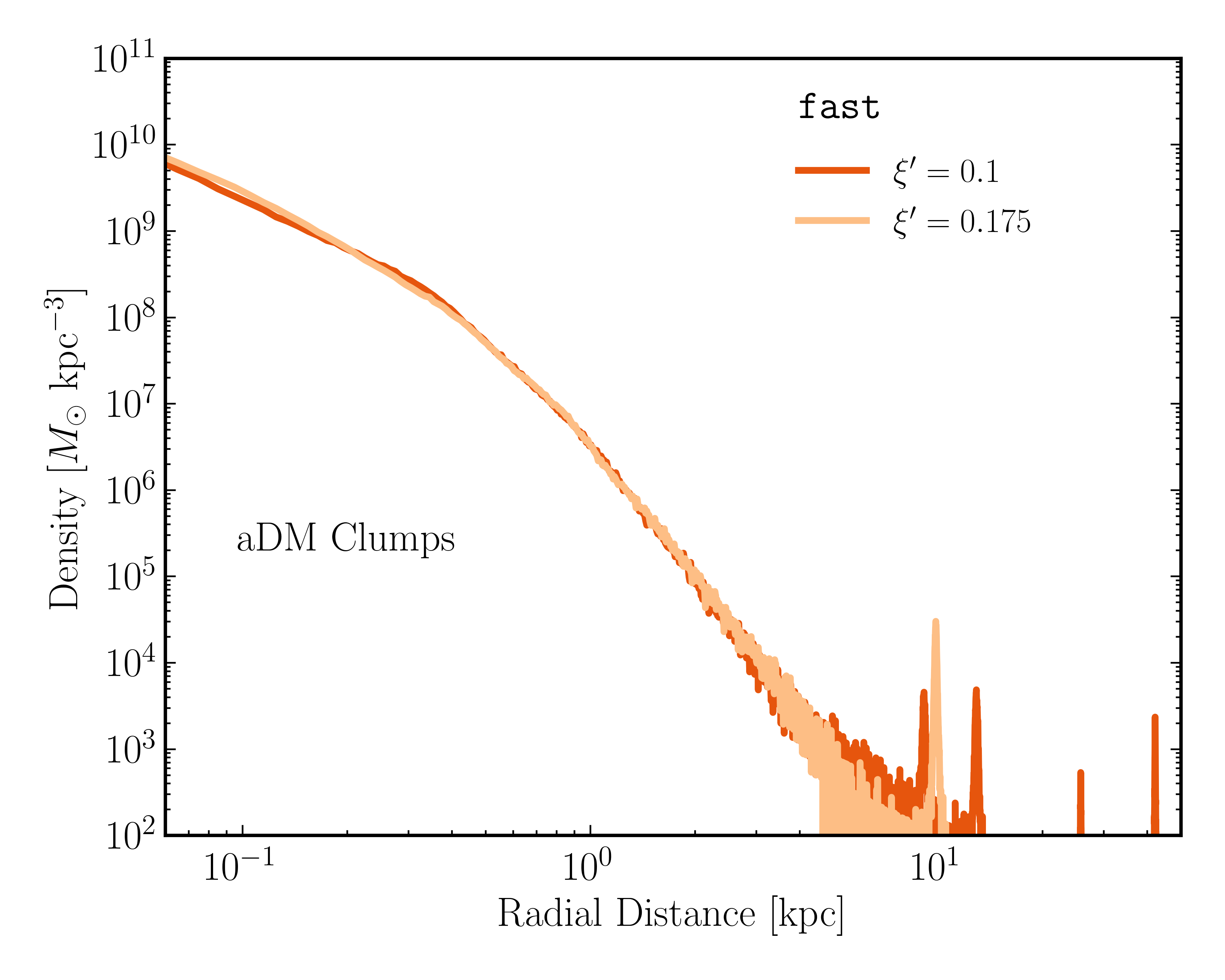}
\includegraphics[trim = {1.5cm, 2cm, 1.5cm, .5cm},width=0.49\textwidth]{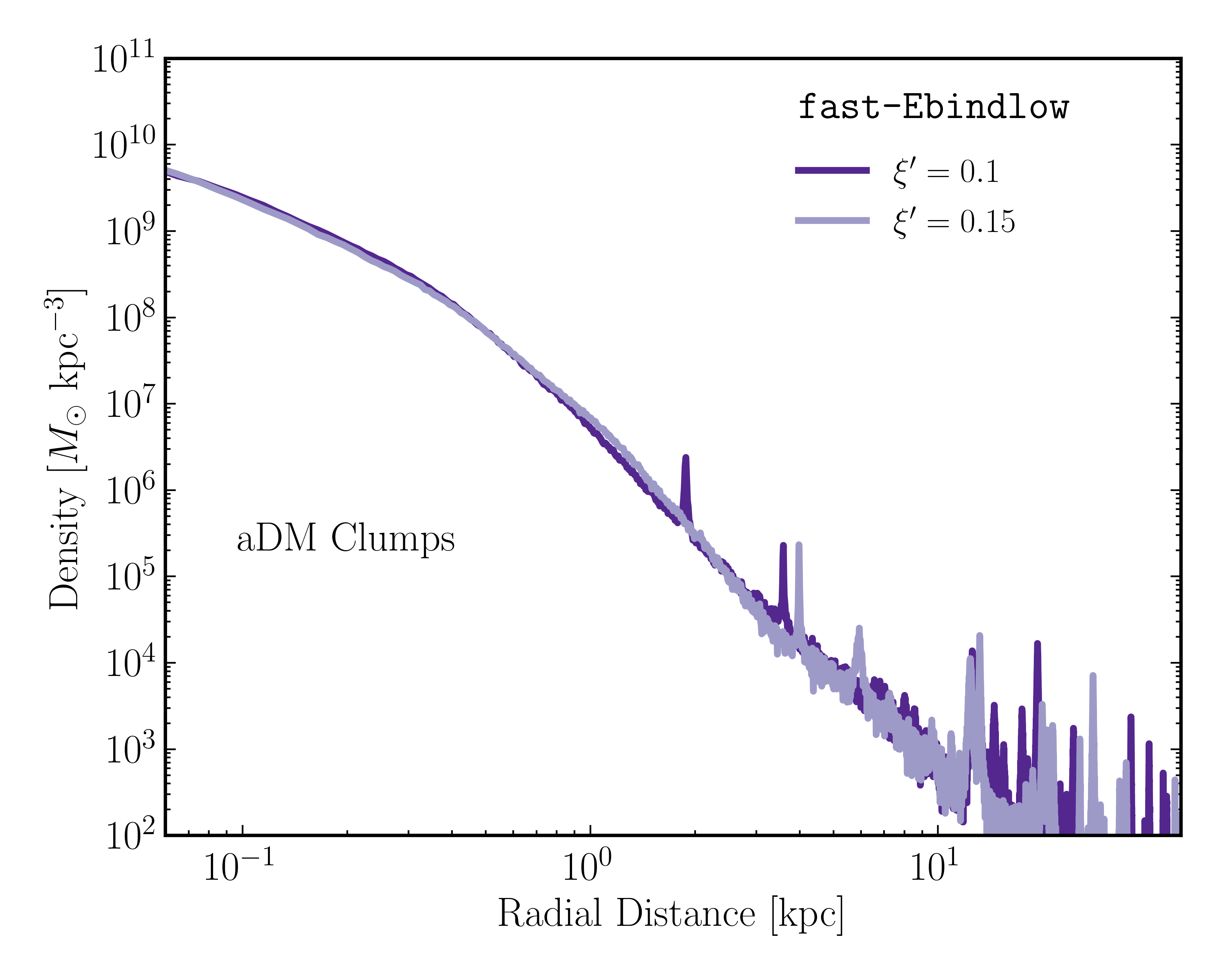}
\includegraphics[trim = {1.5cm, 2cm, 1.5cm, .5cm},width=0.49\textwidth]{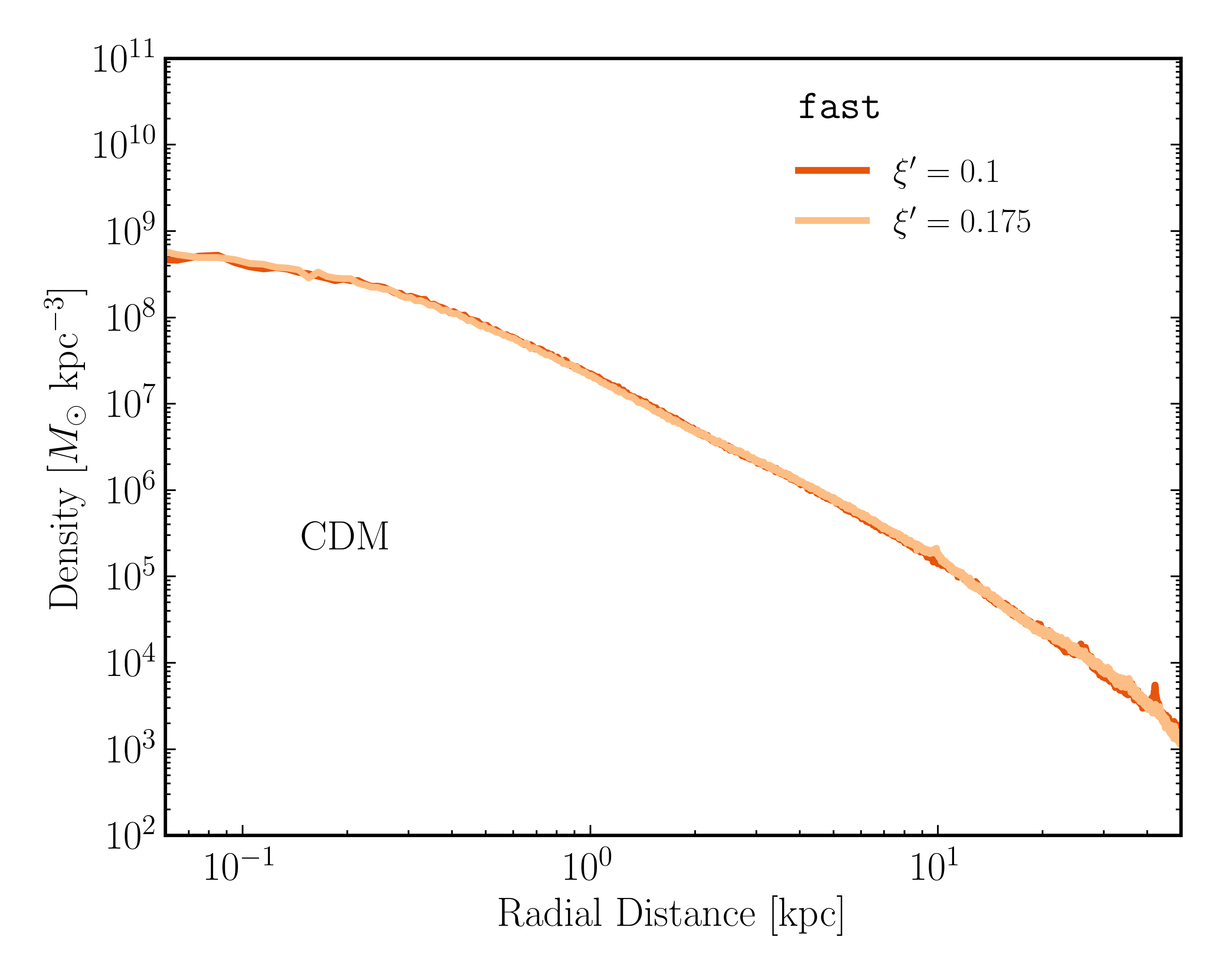}
\hspace{0.2em}
\includegraphics[trim = {1.5cm, 2cm, 1.5cm, .5cm},width=0.49\textwidth]{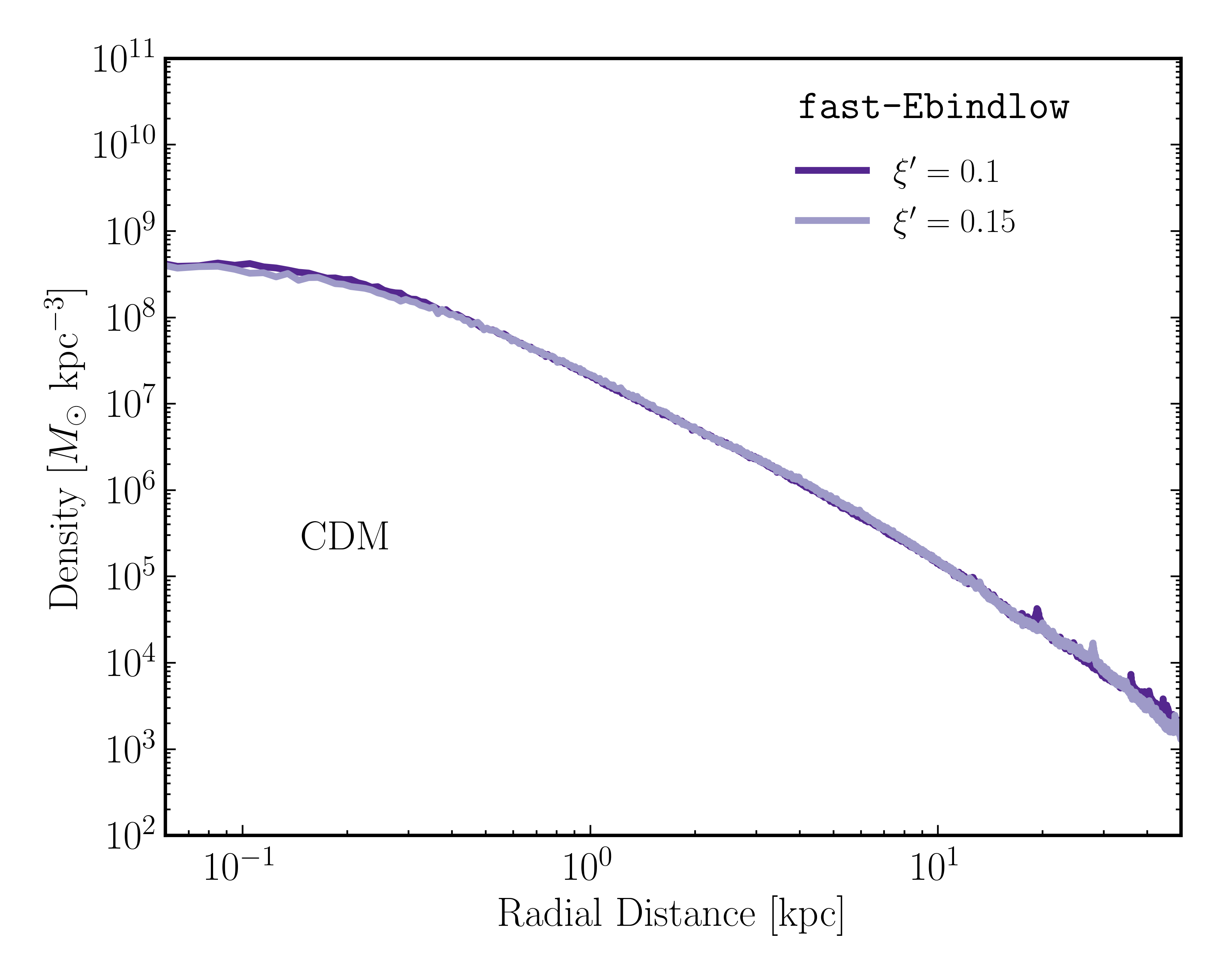}
\caption{
\label{fig:darkcmbratiocomparison} 
\emph{Left column}: Density profile for aDM clumps and CDM, varying over the ratio $\xi = T_{\rm cmb}'/\Tcmb$ and using the $\Bhighthigh$ aDM microphysical parameters. \emph{Right column}: Similar to the left column, but for the $\Blowthigh$ aDM microphysical parameters. Despite differences in the dark CMB temperature, all the density profiles for like particle species agree well. 
}
\end{figure*}

As shown in Fig.~\ref{fig:darkcmbratiocomparison}, 
there are no appreciable differences between the resulting aDM clump and CDM density profiles for the $\Bhighthigh$~(left panel) and $\Blowthigh$~(right panel) simulations. We vary $\xiprime$ from 0.1 up to the largest value allowed by current cosmological constraints for each  parameter point (see \citet{Bansal:2022qbi} for further details). This range of $\xiprime$ corresponds to roughly an order-of-magnitude difference in the effective number of relativistic species in the early Universe ($N_{\rm eff}$). 
Apart from stochastic fluctuations at large radii, the densities agree to within $\lesssim 0.15$ dex at all radii $r\lesssim 10\kpc$ for both aDM clumps and CDM.

\setcounter{equation}{0}
\setcounter{figure}{0} 
\setcounter{table}{0}
\renewcommand{\theequation}{C\arabic{equation}}
\renewcommand{\thefigure}{C\arabic{figure}}
\renewcommand{\thetable}{C\arabic{table}}
\renewcommand*{\theHfigure}{\thefigure}
\renewcommand*{\theHtable}{\thetable}
\renewcommand*{\theHequation}{\theequation}

\clearpage

\section{Numerical Convergence}
\label{sec:sim_convergence}

\newcommand{\trelax}{t_{\rm relax}}
\newcommand{\tcross}{t_{\rm cross}}
\newcommand{\ratioconvergence}{r_{\rm conv.}}
\newcommand{\rhobar}{\overline{\rho}}
\newcommand{\rhocrit}{\rho_{\rm crit}}

A key physical finding of this paper is that the inner aDM clump-dominated regions of the halo thermalise over the course of the halo's evolution and become approximately isothermal by $z=0$.  This section verifies that the thermalisation is not a result of N-body scattering processes. 

To assess simulation convergence, we use the approach from \citet{Power2003} and compare the N-body relaxation timescale, $\trelax$, to the average halo orbit crossing timescale, $\tcross$, at a given galactic radius, $r$. The convergence criterion is as follows:
\begin{equation}
\label{eq:trelaxtcrossratio}
    \frac{\trelax}{\tcross}(<r) = \frac{\sqrt{200}}{8}\frac{N(<r)}{\log(N(<r))}\left(\frac{\rhocrit}{\rhobar(r)}\right)^{1/2} \geq \ratioconvergence \, ,
\end{equation}
where $N(<r)$ is the number of simulation particles within a galactocentric distance $r$, $\rhocrit$ is the critical density of the universe, $\rhobar(r) = 3M(<r)/4\pi r^3$, and $\ratioconvergence$ is a dimensionless constant. 
If the ratio $\trelax/\tcross > \ratioconvergence$, then N-body gravitational particle scattering is not efficient enough to appreciably affect the average densities at that radius, and thus the observed densities are considered numerically resolved.

\citet{Power2003} find convergence with $\ratioconvergence=0.6$ for dark matter-only simulations, but  \citet{Hopkins2018}---who simulate exactly the same halos ($\mq$, $\mv$) and use the same \texttt{GIZMO} version as we do in this paper---find convergence with ratio $\ratioconvergence=0.06$ for CDM-only simulations as well as those with baryons. 

\begin{figure}[h!]\centering
\includegraphics[trim = {0cm, 0cm, 0cm, 0cm},width=0.5\textwidth]{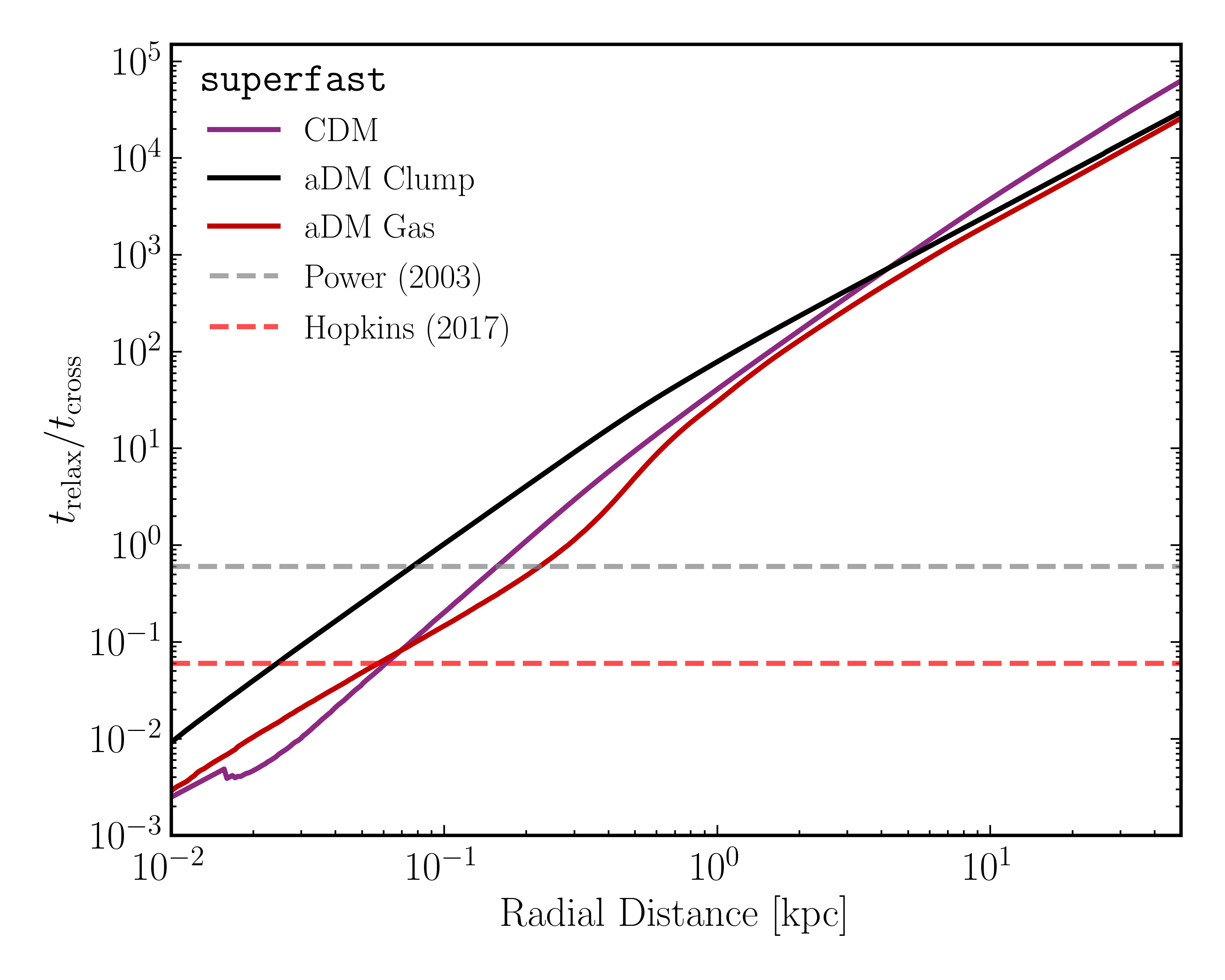}
\vspace{-0.5em}
\caption{
\label{fig:superfast_tcircratio}
Plot of the ratio $t_{\rm relax}/t_{\rm cross}$ (Eq. \ref{eq:trelaxtcrossratio}) from \citet{Power2003} for the fiducial $\Bhightlow$ simulation.  The results are shown separately for CDM~(purple), aDM clumps~(black), and aDM gas~(red). Compared to  \citet{Hopkins2018}, all particle species converge to the fiducial ratio by the radial distance $r\approx 0.06\kpc$, although the aDM clumps converge to this value at much lower radii $r\sim 0.03\kpc$. Compared to the more stringent criteria from \citet{Power2003}, the aDM clumps converge at $r\sim 0.08\kpc$.  
}
\end{figure}

To be conservative, we compute the ratio $\trelax/\tcross$ for the CDM, aDM clump, and aDM gas species separately for $\Bhightlow$ and show the results in Fig.~\ref{fig:superfast_tcircratio}. The aDM clumps achieve the \citet{Hopkins2018} convergence ratio by $r\sim 0.025\kpc$, and the CDM and aDM gas species achieve the ratio by $r\sim 0.06\kpc$. As a result, we conservatively display all plots with $r\gtrsim 0.06\kpc$. Even using the more conservative \citet{Power2003} ratio, the aDM clumps still converge by $r\sim 0.08\kpc$, so our halos should be very well-resolved at the radii where the aDM clumps dominate the inner halo.

\newpage
\section{aDM Simulation Morphological Properties}
\label{sec:morphology_metrics}

\setcounter{equation}{0}
\setcounter{figure}{0} 
\setcounter{table}{0}
\renewcommand{\theequation}{D\arabic{equation}}
\renewcommand{\thefigure}{D\arabic{figure}}
\renewcommand{\thetable}{D\arabic{table}}
\renewcommand*{\theHtable}{\thetable}

\newcommand{\znoughtstar}{z_0'}
\newcommand{\zninetystar}{z_{90}'}

Table~\ref{tab:morphology_metrics_table} lists the morphology metric parameters for aDM clumps and aDM gas in the central halo of all the simulations in this investigation. All the parameters are defined in the main text in Sec. \ref{sec:sim_results} except for $\znoughtstar$ and $\zninetystar$, which are defined in the caption of Tab.~\ref{tab:morphology_metrics_table}.

%

\begin{table}[h!]\centering
\renewcommand{\arraystretch}{1.3}
\begin{tabular}{c|c|ccccccc|c|c}
\Xhline{3\arrayrulewidth} 
\footnotesize
Simulation                          & aDM Particle Type &  $\flatness$    & $\zninety$ & $\rninety$ & $\zhalf$ & $\rhalf$ & $\znoughtstar$ & $\zninetystar$                & $f_{\rm gas}'(0.5 \kpc)$                    & $f_{\rm gas}'(5 \kpc)$ \\ \Xhline{2\arrayrulewidth} 
\multirow{2}{*}{$\Bhightlow$}       & clumps        & 0.63 & 0.27       & 1.1        & 0.078    & 0.31     & 11.1     & \multicolumn{1}{c|}{1.3} & \multicolumn{1}{c|}{\multirow{2}{*}{0.02}}  & \multirow{2}{*}{0.5}   \\
                                    & gas           &  0.91 & 0.14       & 4.1        & 0.038    & 1.2      & -        & \multicolumn{1}{c|}{-}   & \multicolumn{1}{c|}{}                       &                        \\ \hline
\multirow{2}{*}{$\Bhighthigh$}      & clumps        &  0.63 & 0.25       & 0.86       & 0.065    & 0.23     & 10.7     & \multicolumn{1}{c|}{0.6} & \multicolumn{1}{c|}{\multirow{2}{*}{0.09}}  & \multirow{2}{*}{0.6}   \\
                                    & gas           &  0.84 & 0.22       & 3.1        & 0.030    & 0.65     & -        & \multicolumn{1}{c|}{-}   & \multicolumn{1}{c|}{}                       &                        \\ \hline
\multirow{2}{*}{$\Bhightextreme$}   & clumps        &  0.49 & 0.26       & 0.65       & 0.081    & 0.20     & 9.5      & \multicolumn{1}{c|}{2.7} & \multicolumn{1}{c|}{\multirow{2}{*}{0.02}}  & \multirow{2}{*}{0.9}   \\
                                    & gas           & \multicolumn{5}{c}{no gas disk}               & -        & \multicolumn{1}{c|}{-}   & \multicolumn{1}{c|}{}                       &                        \\ \hline
\multirow{2}{*}{$\Bhighthighfhigh$} & clumps        &  0.61 & 0.37       & 1.6        & 0.098    & 0.36     & 10.9     & \multicolumn{1}{c|}{0.8} & \multicolumn{1}{c|}{\multirow{2}{*}{0.03}}  & \multirow{2}{*}{0.4}   \\
                                    & gas           &  0.92 & 0.17       & 4.0        & 0.059    & 1.5      & -        & \multicolumn{1}{c|}{-}   & \multicolumn{1}{c|}{}                       &                        \\ \hline
\multirow{2}{*}{$\Bhightlowmv$}     & clumps        &  0.38 & 0.61       & 1.5        & 0.11     & 0.24     & 7.2      & \multicolumn{1}{c|}{0.3} & \multicolumn{1}{c|}{\multirow{2}{*}{0.04}}  & \multirow{2}{*}{0.2}   \\
                                    & gas           &  0.82 & 0.25       & 5.6        & 0.053    & 0.86     & -        & \multicolumn{1}{c|}{-}   & \multicolumn{1}{c|}{}                       &                        \\ \hline
\multirow{2}{*}{$\Blowthigh$}       & clumps        &  0.65 & 0.31       & 1.4        & 0.084    & 0.35     & 15.2     & \multicolumn{1}{c|}{0.7} & \multicolumn{1}{c|}{\multirow{2}{*}{0.01}}  & \multirow{2}{*}{0.2}   \\
                                    & gas           &  0.88 & 0.11       & 3.2        & 0.035    & 1.4      & -        & \multicolumn{1}{c|}{-}   & \multicolumn{1}{c|}{}                       &                        \\ \hline
\multirow{2}{*}{$\Blowtlow$}        & clumps        &  0.61 & 0.32       & 1.8        & 0.088    & 0.31     & 15.5     & \multicolumn{1}{c|}{0.9} & \multicolumn{1}{c|}{\multirow{2}{*}{0.001}} & \multirow{2}{*}{0.2}   \\
                                    & gas           &  0.92 & 0.17       & 6.6        & 0.070    & 2.91     & -        & \multicolumn{1}{c|}{-}   & \multicolumn{1}{c|}{}                       &                        \\ 
\Xhline{3\arrayrulewidth} 
\end{tabular}
\caption{\label{tab:morphology_metrics_table} Morphology metrics for the aDM clumps and aDM gas in the central halos of all the simulations in this suite. The metrics $\flatness,\, \zninety, \, \rninety, \, \zhalf,\, \rhalf$, and $\fgas$ are all defined in the main text (see Sec.~\ref{sec:sim_results}). The metrics $\znoughtstar$ and $\zninetystar$ focus on the aDM clumps in the region $r\leq 50\kpc$ of the halo at $z=0$.  They respectively correspond to the redshift at which the first of the aDM clumps form and the redshift by which 90\% of these aDM clumps form. 
}
\end{table}


\section{Detailed Density Profiles}
\label{sec:all_density_profiles_and_images}

\setcounter{equation}{0}
\setcounter{figure}{0} 
\setcounter{table}{0}
\renewcommand{\theequation}{E\arabic{equation}}
\renewcommand{\thefigure}{E\arabic{figure}}
\renewcommand{\thetable}{E\arabic{table}}
\renewcommand*{\theHfigure}{\thefigure}
\renewcommand*{\theHtable}{\thetable}
\renewcommand*{\theHequation}{\theequation}

\begin{figure}[t!]\centering
\includegraphics[trim = {0cm, 0cm, 42cm, 0cm},width=0.9\textwidth]{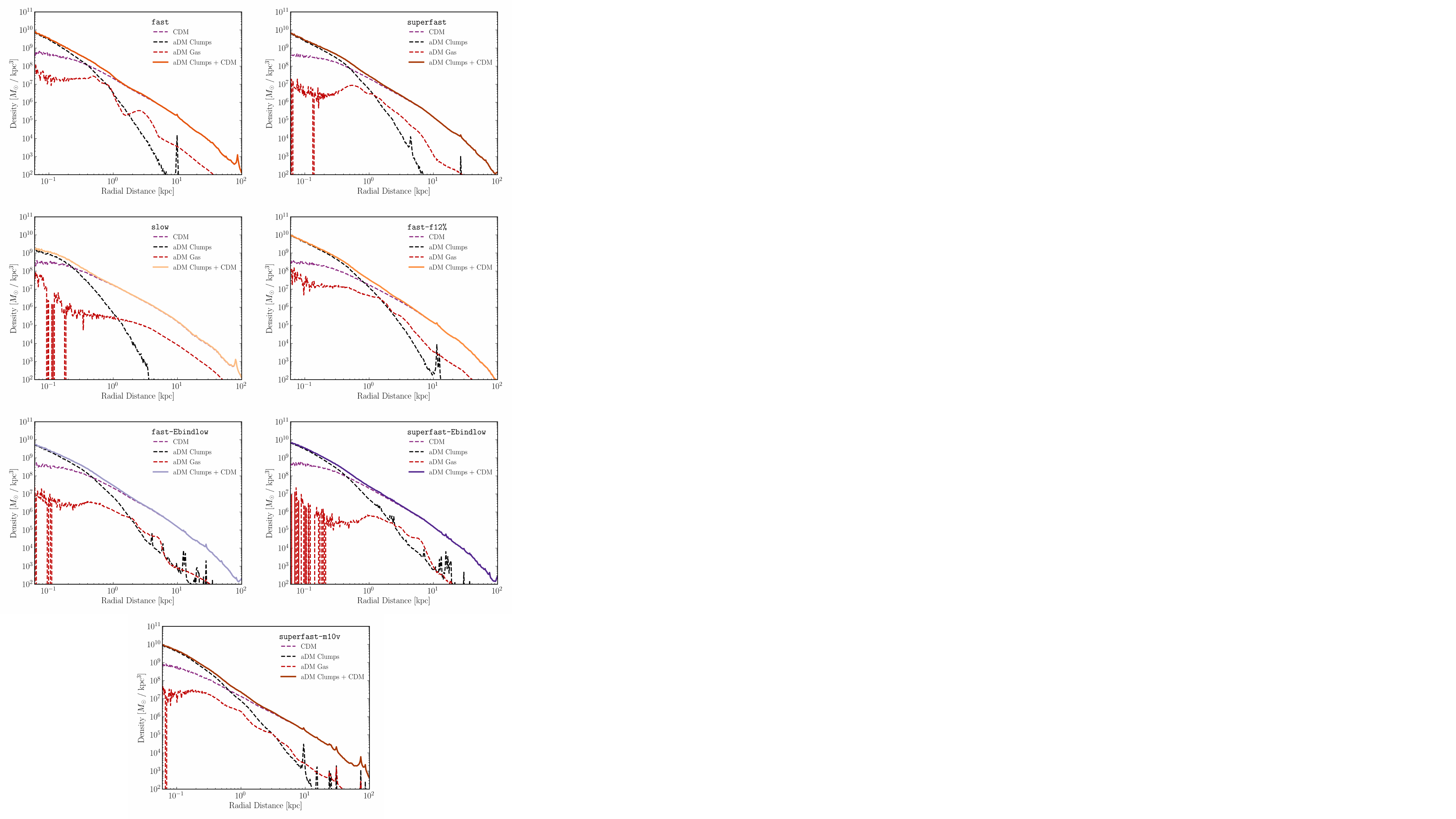}
\caption{
\label{fig:all_density_profiles}
The CDM, aDM clump, and aDM gas density profiles for all simulations in the suite.  
}
\end{figure}

\begin{figure}[t!]\centering
\includegraphics[trim = {0cm, 1cm, 0cm, 0cm},width=0.5\textwidth]{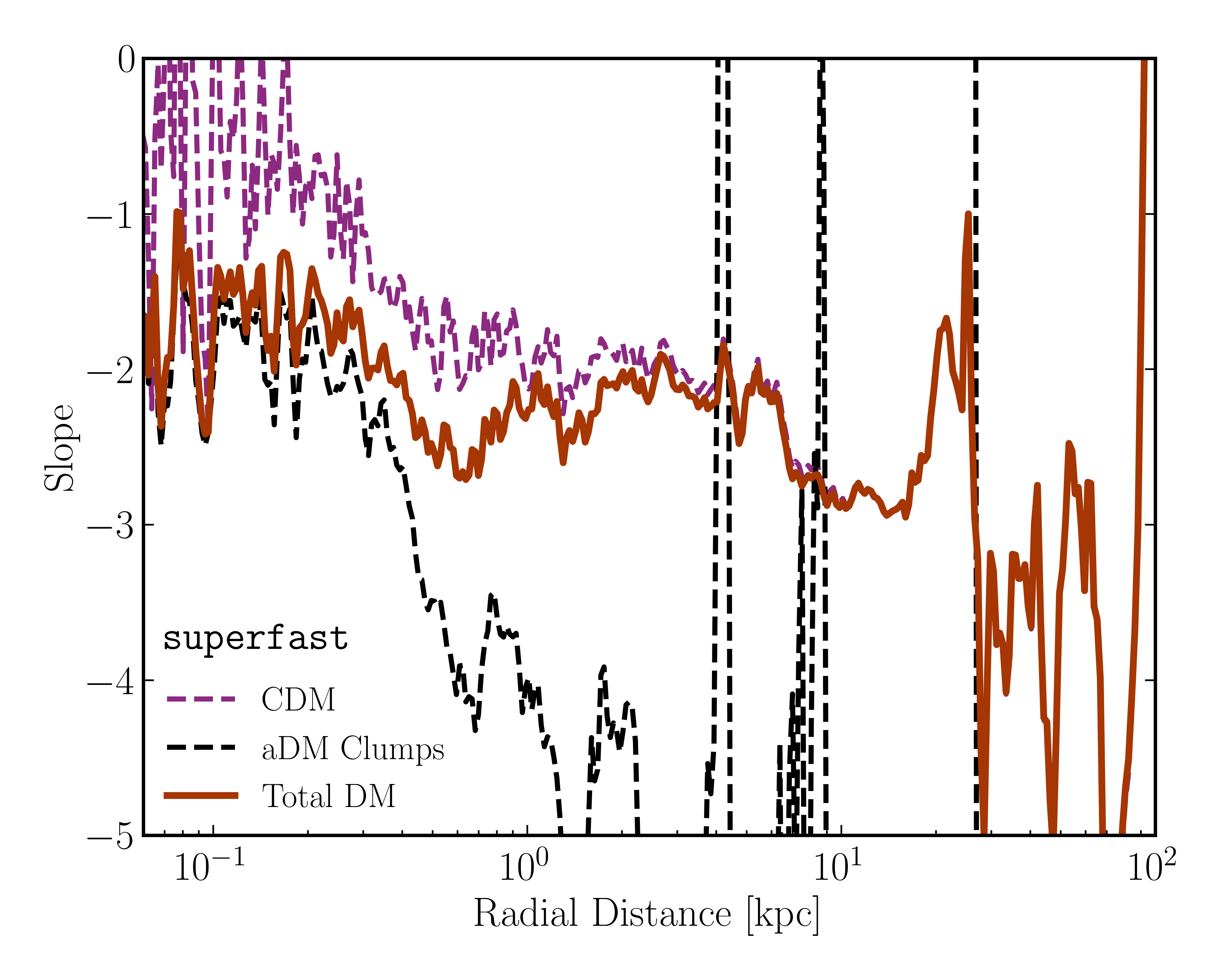}
\caption{
\label{fig:fiducial_slope_plot}
The slopes of CDM, aDM clumps and all the dark matter~(DM), including aDM gas. 
}
\end{figure}

This section presents the density profiles for all of the simulations in the suite, breaking them down by species (CDM, aDM clumps, aDM gas). These density profiles are showcased in Fig.~\ref{fig:all_density_profiles}. We also present the density profile slopes for the fiducial simulation $\Bhightlow$ in Fig.~\ref{fig:fiducial_slope_plot}.

\clearpage

\section{Fitting Function Paramaters}
\label{sec:fitting_parameters_forall}

\setcounter{equation}{0}
\setcounter{figure}{0} 
\setcounter{table}{0}
\renewcommand{\theequation}{F\arabic{equation}}
\renewcommand{\thefigure}{F\arabic{figure}}
\renewcommand{\thetable}{F\arabic{table}}
\renewcommand*{\theHfigure}{\thefigure}
\renewcommand*{\theHtable}{\thetable}
\renewcommand*{\theHequation}{\theequation}

This appendix presents the aDM clump and CDM fitting function parameters (from Eqs. \ref{eq:rhocdm} and \ref{eq:rhoadm}) for the simulations in the dwarf suite. To conduct the fit, we use least-squares regression on the simulated aDM clump densities in the region $0.06\kpc \leq r \leq 5\kpc$ and the same  for CDM densities in the region $0.06\kpc\leq r \leq 50\kpc$. At greater galactocentric distances than $5\kpc$, the simulated aDM clump densities are statistically limited, so we ignore those regions for the aDM fit.

\begin{table}[h!]
\footnotesize
\renewcommand{\arraystretch}{1.5}
\begin{tabular}{c|c|cccc}
  \Xhline{3\arrayrulewidth} 
  \multirow{4}{*}{$\Bhightlow$} & \multirow{2}{*}{\shortstack{aDM\\clumps}} & $\rho_0'\,[\msun\kpc^{-3}]$ & $\rsprime\,[\kpc]$ & ${\gamma}$ &  ${\beta}$\\
\cline{3-6}
 & & $(1.14\pm 0.13)\times 10^8$ & $0.65\pm 0.02$ & $1.67 \pm 0.03$ & $5.55\pm 0.06$  \\
 \cline{2-6}
 & \multirow{2}{*}{{CDM}} & $\rho_0\,[\msun\kpc^{-3}]$ & $r_s\,[\kpc]$ & ${r_c}\,[\kpc]$ &  ${\alpha_E}$\\
 \cline{3-6}
 & & $(7.6\pm 4.8)\times 10^7$ & $0.66\pm 0.2$ & $0.18\pm 0.02$ & $0.096 \pm 0.01$\\
  \Xhline{2\arrayrulewidth}
  \multirow{4}{*}{$\Bhighthigh$} & \multirow{2}{*}{\shortstack{aDM\\clumps}} & $\rho_0'\,[\msun\kpc^{-3}]$ & $\rsprime\,[\kpc]$ & ${\gamma}$ &  ${\beta}$\\
\cline{3-6}
 & & $(1.6\pm 0.1)\times 10^8$ & $0.52\pm 0.01$ & $1.7 \pm 0.01$ & $5.4\pm 0.05$  \\
 \cline{2-6}
 & \multirow{2}{*}{{CDM}} & $\rho_0\,[\msun\kpc^{-3}]$ & $r_s\,[\kpc]$ & ${r_c}\,[\kpc]$ &  ${\alpha_E}$\\
 \cline{3-6}
 & & $(1.0\pm 0.7)\times 10^8$ & $0.55\pm 0.2$ & $0.15\pm 0.02$ & $0.082 \pm 0.01$\\
 \Xhline{2\arrayrulewidth}
  \multirow{4}{*}{$\Bhightextreme$} & \multirow{2}{*}{\shortstack{aDM\\clumps}} & $\rho_0'\,[\msun\kpc^{-3}]$ & $\rsprime\,[\kpc]$ & ${\gamma}$ &  ${\beta}$\\
\cline{3-6}
 & & $(5.4\pm 0.8)\times 10^8$ & $0.30\pm 0.02$ & $0.54 \pm 0.06$ & $5.7\pm 0.1$  \\
 \cline{2-6}
 & \multirow{2}{*}{{CDM}} & $\rho_0\,[\msun\kpc^{-3}]$ & $r_s\,[\kpc]$ & ${r_c}\,[\kpc]$ &  ${\alpha_E}$\\
 \cline{3-6}
 & & $(2.7\pm 1.8)\times 10^7$ & $1.0\pm 0.3$ & $0.18\pm 0.03$ & $0.094 \pm 0.012$\\
  \Xhline{2\arrayrulewidth}
  \multirow{4}{*}{$\Bhightlowmv$} & \multirow{2}{*}{\shortstack{aDM\\clumps}} & $\rho_0'\,[\msun\kpc^{-3}]$ & $\rsprime\,[\kpc]$ & ${\gamma}$ &  ${\beta}$\\
\cline{3-6}
 & & $(9.5\pm 1.6)\times 10^8$ & $0.27\pm 0.02$ & $1.6 \pm 0.1$ & $3.73\pm 0.02$  \\
 \cline{2-6}
 & \multirow{2}{*}{{CDM}} & $\rho_0\,[\msun\kpc^{-3}]$ & $r_s\,[\kpc]$ & ${r_c}\,[\kpc]$ &  ${\alpha_E}$\\
 \cline{3-6}
 & & $(5.4\pm 3.2)\times 10^6$ & $1.8\pm 0.5$ & $0.07\pm 0.01$ & $0.051 \pm 0.01$\\
   \Xhline{2\arrayrulewidth}
  \multirow{4}{*}{$\Bhighthighfhigh$} & \multirow{2}{*}{\shortstack{aDM\\clumps}} & $\rho_0'\,[\msun\kpc^{-3}]$ & $\rsprime\,[\kpc]$ & ${\gamma}$ &  ${\beta}$\\
\cline{3-6}
 & & $(2.2\pm 0.2)\times 10^8$ & $0.59\pm 0.02$ & $1.67 \pm 0.03$ & $4.49\pm 0.04$  \\
 \cline{2-6}
 & \multirow{2}{*}{{CDM}} & $\rho_0\,[\msun\kpc^{-3}]$ & $r_s\,[\kpc]$ & ${r_c}\,[\kpc]$ &  ${\alpha_E}$\\
 \cline{3-6}
 & & $(2.9\pm 1.3)\times 10^7$ & $0.95\pm 0.18$ & $0.18\pm 0.02$ & $0.10 \pm 0.01$\\
 \Xhline{2\arrayrulewidth}
  \multirow{4}{*}{$\Blowtlow$} & \multirow{2}{*}{\shortstack{aDM\\clumps}} & $\rho_0'\,[\msun\kpc^{-3}]$ & $\rsprime\,[\kpc]$ & ${\gamma}$ &  ${\beta}$\\
\cline{3-6}
 & & $(2.5\pm 1.0)\times 10^8$ & $0.43\pm 0.06$ & $1.75 \pm 0.12$ & $4.13\pm 0.10$  \\
 \cline{2-6}
 & \multirow{2}{*}{{CDM}} & $\rho_0\,[\msun\kpc^{-3}]$ & $r_s\,[\kpc]$ & ${r_c}\,[\kpc]$ &  ${\alpha_E}$\\
 \cline{3-6}
 & & $(1.1\pm 0.6)\times 10^8$ & $0.54\pm 0.14$ & $0.16\pm 0.01$ & $0.084 \pm 0.007$\\
  \Xhline{2\arrayrulewidth}
  \multirow{4}{*}{$\Blowthigh$} & \multirow{2}{*}{\shortstack{aDM\\clumps}} & $\rho_0'\,[\msun\kpc^{-3}]$ & $\rsprime\,[\kpc]$ & ${\gamma}$ &  ${\beta}$\\
\cline{3-6}
 & & $(1.2\pm 0.2)\times 10^8$ & $0.61\pm 0.03$ & $1.65 \pm 0.04$ & $4.88\pm 0.07$  \\
 \cline{2-6}
 & \multirow{2}{*}{{CDM}} & $\rho_0\,[\msun\kpc^{-3}]$ & $r_s\,[\kpc]$ & ${r_c}\,[\kpc]$ &  ${\alpha_E}$\\
 \cline{3-6}
 & & $(2.7\pm 2.4)\times 10^8$ & $0.37\pm 0.14$ & $0.21\pm 0.02$ & $0.076 \pm 0.009$\\
  \Xhline{3\arrayrulewidth}
\end{tabular}
\caption{\label{tab:fitting_parameters_forall} Fitting function parameters (Eqs. \ref{eq:rhocdm} \& \ref{eq:rhoadm}) for CDM and aDM clumps for all simulations in the suite. The error bars correspond to the $1\sigma$ fit uncertainties derived from the least-squares fitting process.
}
\end{table}

\end{document}